\documentclass[aps,prb,reprint]{revtex4-2}

\usepackage{graphicx}
\usepackage{dcolumn}
\usepackage{bm}
\usepackage{amsmath}
\usepackage{amssymb}
\usepackage{float}
\usepackage{multirow}
\usepackage{mathrsfs}
\usepackage{xcolor}
\usepackage{comment}
\usepackage{mathtools}
\usepackage{anyfontsize}
\usepackage{textcomp}
\usepackage[caption=false]{subfig}

\DeclareUnicodeCharacter{2212}{\textminus}

\begin{document}

\title{Thermodynamics and Geometrical Optics of Reissner–Nordström–de Sitter Black Holes in noncommutative Geometry}

\author{Phongsakorn Sereewat}
 \email{phongsakorn.ser@student.mahidol.edu}
\altaffiliation{Department of Physics, Faculty of Science, Mahidol University, 272 Rama VI Road, Ratchathewi, Bangkok 10400, Thailand}
 
\author{David Senjaya}%
 \email{davidsenjaya@protonmail.com}
\affiliation{Department of Physics, Faculty of Science, Mahidol University, 272 Rama VI Road, Ratchathewi, Bangkok 10400, Thailand}

\author{Piyabut Burikham}
 \email{piyabut@gmail.com}
\affiliation{High Energy Physics Theory Group, Department of Physics, Faculty of Science,
Chulalongkorn University, Bangkok 10330, Thailand}

\date{\today}

\begin{abstract}
We investigate the thermodynamic, optical, and dynamical properties of Reissner-Nordström-de Sitter black holes in a noncommutative spacetime endowed with a fundamental minimal length scale $\Theta$. Working within a consistent two-horizon framework, we formulate an effective first law of thermodynamics and introduce an entanglement entropy that captures correlation effects between the event and cosmological horizons. Imposing the lukewarm condition, under which both horizons share a common temperature, uniquely fixes the entropy correction and yields closed-form expressions for the temperature, pressure, and electric potential. An analysis of the heat capacity, entropy, Helmholtz free energy, and pressure reveals a noncommutativity-induced second-order phase transition, highlighting the crucial role of short-distance structure in the critical behavior of the system. On the optical side, we examine photon motion and weak gravitational lensing in this geometry. We show that noncommutativity deforms the photon effective potential and shifts the critical impact parameter separating capture from scattering. Using the Gauss-Bonnet method, we derive an analytical expression for the weak deflection angle and demonstrate how charge and the minimal length scale modify the optical response in the presence of a cosmological constant. Finally, we establish a direct link between geometry and dynamics by analyzing the Lyapunov exponent of circular photon orbits and the imaginary part of the quasinormal mode frequencies. We show that the black hole parameters leave distinct and systematic imprints on both orbital instability and damping: increasing the mass suppresses the instability and leads to longer-lived modes, charge enhances both the divergence of null geodesics and the decay rate, the cosmological constant slightly reduces damping, and noncommutativity amplifies the instability and accelerates relaxation.
\end{abstract}

\maketitle

\section{Introduction}
Black holes are among the most striking predictions of general relativity~(GR). Despite arising from the highly nonlinear Einstein equations, they exhibit a remarkable degree of simplicity: at the macroscopic level, a black hole is completely specified by only a few parameters, namely its mass $M$, electric charge $Q$ and angular momentum $J$ \cite{Israel:1967wq, Carter:1971zc, Bekenstein:1972tm, Bekenstein:1973ur, Izmailov:2019cqr}. This simplicity laid the foundation for black hole thermodynamics, where black holes are treated as genuine thermodynamic systems possessing entropy and temperature, following Hawking's discovery of black hole radiation \cite{Hawking:1976de}. These insights culminated in the formulation of the four laws of black hole thermodynamics, revealing an unexpected and profound link between gravity, thermodynamics and quantum physics.

A major advance in this subject came with the realization that the cosmological constant $\Lambda$ can be incorporated into black hole thermodynamics as a dynamical quantity \cite{Dolan:2010ha,Kubiznak:2016qmn}. In this extended framework, $\Lambda$ is interpreted as a thermodynamic pressure, while its conjugate variable plays the role of thermodynamic volume. This perspective, known as black hole chemistry \cite{Mann:2024sru,Mann:2025xrb}, leads to a reinterpretation of the black hole mass as enthalpy rather than internal energy. One of the most intriguing outcomes of this approach is the emergence of phase behavior closely resembling that of ordinary thermodynamic systems. Charged and rotating AdS black holes, for instance, display Van der Waals type phase transitions \cite{Chamblin:1999hg,Kubiznak:2012wp,Wei:2015ana,Cheng:2016bpx}, highlighting the rich thermodynamic structure hidden within classical gravitational solutions.

Alongside these developments, noncommutative geometry has emerged as a compelling framework for extending the classical description of spacetime, motivated by ideas from quantum gravity \cite{Nicolini:2008aj}. By introducing a fundamental length scale, noncommutative spacetime effectively smooths out short-distance singularities and leads to modified black hole solutions. In this approach, point-like mass and charge distributions are replaced by smeared profiles, commonly modeled by Gaussian \cite{Tejeiro:2010gu,Gingrich:2010ed} or Lorentzian functions \cite{Liang:2012vx, Hamil:2024ppj}. As a result, the thermodynamic and geometric properties of black holes are significantly altered.

One of the most notable consequences of non-commutativity is the regularization of black hole thermodynamics at small length scales. In particular, divergences in the Hawking temperature are removed and a minimum black hole mass naturally emerges, determined by the fundamental length scale. This minimum mass acts as a remnant, preventing complete evaporation and providing a possible resolution to the end-point problem of black hole evaporation \cite{Nasseri:2005ji}. Such effects are especially relevant for mini black holes, where spacetime fuzziness induced by the uncertainty principle becomes dominant \cite{Campos:2021sff}.

Noncommutative corrections have been shown to influence a wide range of thermodynamic quantities, including entropy, heat capacity and phase structure. These corrections typically modify the entropy beyond the standard area law and significantly affect the heat capacity, leading to qualitative changes in black hole stability \cite{Campos:2021sff}. Extensions of this framework have also considered the presence of additional matter fields, such as quintessence, which further modify the black hole geometry and thermodynamic behavior. Within this broader context, Reissner-Nordstrom black hole solutions incorporating noncommutative geometry effects have been extensively studied with particular emphasis on the role of spacetime fuzziness in shaping black hole entropy and thermodynamic stability (see \cite{DimitrijevicCiric:2024ibc,Hamil:2025vey,ElHadri:2025jof} and references therein).

In this work, we explore the thermodynamics of the noncommutative Reissner-Nordstrom black hole within the framework of effective black hole thermodynamics \cite{Zhang:2016nws,Guo:2025evx}. We systematically analyze the thermodynamic quantities derived from the mass function, effective temperature, entropy, electric potential and thermodynamic volume, as well as response functions such as the heat capacity and compressibility. The stability and phase structure of the system are examined through the behavior of the entropy and Helmholtz free energies.


\section{RN-dS Black Holes in noncommutative Space}

In this section, we will review Reissner-Nordstrom-de Sitter  black hole solution in a noncommutative spacetime background \cite{Hamil_2025}. The spacetime is characterized by a positive cosmological constant, $\Lambda>0$ and geometry is asymptotically de Sitter. Our objective is to understand how the presence of a fundamental minimal length scale, arising from spacetime non-commutativity, modifies the classical charged black hole solution.

Let us start with the Einstein-Maxwell theory in four dimensions \cite{Wald1984,Bardeen1973}
\begin{equation}
\label{action}
S=\int \left(R+2\Lambda+2 F_{\mu\nu}F^{\mu\nu}
\right)\sqrt{-g}d^4x,
\end{equation}
where $R$ denotes the Ricci scalar and $ F_{\mu\nu}=\partial_\mu A_\nu-\partial_\nu A_\mu$ is the electromagnetic field strength tensor. Throughout this work, we restrict attention to the standard Maxwell field and do not introduce phantom or exotic matter contributions.

Varying the action with respect to the metric and the gauge potential yields the Einstein and Maxwell equations,
\begin{gather}
R_{\mu\nu}-\frac{1}{2}g_{\mu\nu}R+\Lambda g_{\mu\nu}=8\pi\left(
T_{\mu\nu}^{matter}+T_{\mu\nu}^{EM}\right),\label{EinsteinEq}\\
T^{\rm EM}_{\mu\nu} =\frac{1}{4\pi}\left( F{\mu\lambda}F_{\nu}{}^{\lambda} -\frac{1}{4}g_{\mu\nu}F_{\alpha\beta}F^{\alpha\beta} \right) \label{TEM},\\
\frac{1}{\sqrt{-g}}\partial_\mu\left(\sqrt{-g}F^{\mu\nu}\right)
=4\pi J^\nu .\label{MaxwellEq}
\end{gather}
which couple the spacetime geometry to the matter and electromagnetic sources.

A key ingredient of the noncommutative framework is the replacement of point-like sources by smeared distributions. Physically, this reflects the existence of a minimal length scale $\sqrt{\Theta }$, which prevents arbitrarily sharp localization of matter and charge. In this work, we adopt Lorentzian profiles for the mass and electric charge densities \cite{ElHadriJemri2025},
\begin{align}
\rho_{\text{matt}}(r,\Theta )&=\frac{M\sqrt{\Theta }}{\pi^{3/2}(r^2 +\pi\Theta )^2 },\label{LorentzMass}\\
\rho_{\text{em}}(r,\Theta )&=\frac{Q\sqrt{\Theta }}{\pi^{3/2}\left(r^2 +\pi\Theta \right)^2 }, \label{LorentzCharge}
\end{align}
where $M$ and $e \equiv 4\pi Q $ denote the total black hole mass and electric charge of the black hole, respectively. These smeared sources reduce to the usual Dirac delta distributions in the commutative limit $\Theta \to0$.

We consider a static and spherically symmetric spacetime, whose line element can be written in the form
\begin{equation}
ds^2 =- f(r)dt^2 +\frac{dr^2 }{f(r)}+r^2 \left(d\theta ^2 +\sin^2 \theta  d\phi^2 \right).
\label{metric}
\end{equation}

Within this metric ansatz, all geometric properties of the spacetime are fully encoded in the single metric function $f(r)$, which we parametrize as
\begin{equation}
f(r)=1-\frac{\Lambda}{3}r^2 -F(r,\Theta ),
\end{equation}
where $\Lambda$ denotes the cosmological constant and the function $F(r,\Theta )$ encodes the contributions arising from the noncommutative matter and electromagnetic sources, with $\Theta $ characterizing the non-commutativity scale.

Due to spherical symmetry and time independence, the Maxwell field strength admits only one independent nonvanishing component. Consequently, the electromagnetic field tensor takes the form
\begin{equation}
F_{tr}=-F_{rt}=-E(r),
\end{equation}
where $E(r)$ represents the radial electric field.

Raising the indices with the inverse metric associated with \eqref{metric}, one finds
\begin{equation}
F^{rt}=g^{rr}g^{tt}F_{tr}=E(r),
\label{FE}
\end{equation}
which shows that the contravariant field component coincides with the electric field.

Setting $\nu = 0$ in the Maxwell equation \eqref{MaxwellEq} and using the metric determinant $\sqrt{-g}=r^2 \sin\Theta $ together with a static and spherically symmetric charge distribution,
\begin{equation}
J^\mu=\big(\rho_{\rm em}(r),0,0,0\big),
\label{eq:Jmu}
\end{equation}
as well as the relation \eqref{FE}, one readily obtains Gauss’s law in curved spacetime,
\begin{equation}
\frac{1}{r^2 }\frac{d}{dr}\left(r^2 E(r)\right) = 4\pi\rho_{\mathrm{el}}(r).
\label{GaussLaw}
\end{equation}

Upon substituting the Lorentzian noncommutative charge density \eqref{LorentzCharge} into \eqref{GaussLaw}, the equation takes the form
\begin{equation}
\frac{d}{dr}\left(r^2 E(r)\right)=\frac{4Q\sqrt{\Theta }}{\sqrt{\pi}}\frac{r^2 }{\left(r^2 +\pi\Theta \right)^2 }.
\label{GaussInt1}
\end{equation}
Integrating, one obtains
\begin{equation}
r^2 E(r)=\frac{4Q\sqrt{\Theta }}{\sqrt{\pi}}\int\frac{r^2 }{\left(r^2 +\pi\Theta \right)^2 }dr,
\label{GaussInt2}
\end{equation}
where the integration constant has been fixed by imposing the regularity condition $E(0)=0$.

The integrand may be decomposed as
\begin{equation}
\frac{r^2 }{\left(r^2 +\pi\Theta \right)^2 }
=\frac{1}{r^2 +\pi\Theta }
-\frac{\pi\Theta }{\left(r^2 +\pi\Theta \right)^2 },
\end{equation}
which allows the integral to be evaluated analytically. The relevant primitives are
\begin{align}
\int\frac{dr}{r^2 +\pi\Theta }
&=\frac{\arctan\left(\frac{r}{\sqrt{\pi\Theta }}\right)}{\sqrt{\pi\Theta }},\\
\int\frac{dr}{\left(r^2 +\pi\Theta \right)^2 }
&=\frac{r}{2\pi\Theta \left(r^2 +\pi\Theta \right)}
+\frac{\arctan\left(\frac{r}{\sqrt{\pi\Theta }}\right)}{2(\pi\Theta )^{3/2}}
.
\end{align}

Combining these results, one finds
\begin{equation}
\int\frac{r^2 }{\left(r^2 +\pi\Theta \right)^2 }dr
=\frac{\arctan\left(\frac{r}{\sqrt{\pi\Theta }}\right)}{2\sqrt{\pi\Theta }}-\frac{r}{2\left(r^2 +\pi\Theta \right)}.
\end{equation}

Substituting this expression into \eqref{GaussInt2} and dividing by $r^2 $, the radial electric field is finally obtained as
\begin{equation}
E(r)=\frac{2Q}{\pi r^2 }\left[\arctan\left(\frac{r}{\sqrt{\pi\Theta }}\right)-\frac{r\sqrt{\pi\Theta }}{r^2 +\pi\Theta }\right].
\label{ElectricField}
\end{equation}

We note that the electric field remains finite at the origin and asymptotically approaches the Coulomb form $E(r)=Q/r^2 $ in the commutative limit $\Theta \to 0$. Moreover, the Maxwell invariant reads
\begin{equation}
F_{\alpha\beta}F^{\alpha\beta}=2F_{tr}F^{tr}=-2E(r)^2 ,
\end{equation}
and one further finds
\begin{equation}
F_{t\lambda}F_{t}{}^{\lambda}=F_{tr}g^{rr}F_{tr}=f(r)E(r)^2 .
\end{equation}

Substituting these expressions into \eqref{TEM}, the $tt$ component of the electromagnetic stress-energy tensor becomes
\begin{align}
T^{\rm EM}_{tt} &=\frac{1}{4\pi} \left[ f(r)E(r)^2 -\frac14 g_{tt}\left(-2E(r)^2\right) \right] \nonumber\\
&=\frac{f(r)E(r)^2}{8\pi}.
\end{align}
Raising an index with $g^{tt}=-1/f(r)$, we obtain
\begin{equation}
T^{t}{}_{t}=-\frac{E(r)^2}{8\pi}.
\end{equation}

For the metric \eqref{metric}, the $(t,t)$ component of the Einstein tensor is given by
\begin{equation}
G^{t}{}_{t}
= -\Lambda
- \frac{1}{r^2 }\frac{d}{dr}\left[rF(r,\Theta )\right].
\end{equation}

Substituting this expression into the Einstein equations \eqref{EinsteinEq}, one obtains
\begin{equation}
\frac{1}{r^2 } \frac{d}{dr} \left(rF(r,\Theta )\right)
= 8\pi\rho_{\rm {total}}(r),
\label{Feq}
\end{equation}
where the total energy density is defined as
\begin{gather}
\rho_{\rm total}(r) = \rho_{\rm matt}(r) + \rho_{\rm em}(r),\\ \quad \rho_{\rm em}(r) \equiv -T^{t}{}_{t} = \frac{E(r)^2 }{8\pi}.
\end{gather}

Equation \eqref{Feq} can be integrated to yield
\begin{equation}
F(r,\Theta ) =\frac{1}{r}\int_{0}^{r} \left[ 8\pi r'^2 \rho_{\rm matt}(r',\Theta ) +r'^2 E(r')^2  \right]dr' +\frac{c}{r}, \label{Fgeneral}
\end{equation}
where $c$ is an integration constant.

Substituting the Lorentzian mass density \eqref{LorentzMass} and the radial electric field \eqref{ElectricField} into \eqref{Fgeneral}, the metric function can be expressed as
\begin{equation}
f(r)
=1-\frac{\Lambda}{3}r^2 -\frac{c}{r}
- F_{\text{matt}}(r,\Theta )
- F_{\text{em}}(r,\Theta ),
\end{equation}
with
\begin{align}
F_{\rm matt}(r,\Theta ) &=\frac{4M}{\sqrt{\pi}r} \left[ \arctan\left(\frac{r}{\sqrt{\pi\Theta }}\right) -\frac{r\sqrt{\pi\Theta }}{r^2+\pi\Theta } \right], \\ 
F_{\rm em}(r,\Theta ) &=\frac{Q^2}{r^2} -\frac{4Q^2}{\pi r^2} \left[ \arctan\left(\frac{r}{\sqrt{\pi\Theta }}\right) -\frac{r\sqrt{\pi\Theta }}{r^2+\pi\Theta } \right]^2.
\end{align}

Substituting these expressions into the Einstein equations and performing a single integration, the exact metric function can be written as
\begin{multline}
f(r) =1-\frac{\Lambda}{3}r^2-\frac{c}{r} -\frac{4M}{\sqrt{\pi}r} \left[ \arctan\left(\frac{r}{\sqrt{\pi\Theta }}\right) -\frac{r\sqrt{\pi\Theta }}{r^2+\pi\Theta } \right] \\+\frac{Q^2}{r^2} -\frac{4Q^2}{\pi r^2} \left[ \arctan\left(\frac{r}{\sqrt{\pi\Theta }}\right) -\frac{r\sqrt{\pi\Theta }}{r^2+\pi\Theta } \right]^2 . \label{metric_exact}
\end{multline}
where $c$ is an integration constant.

For small noncommutative parameter $\Theta $ and fixed $r>0$, the relevant asymptotic expansions are
\begin{align}
\arctan\left(\frac{r}{\sqrt{\pi\Theta }}\right) &= \frac{\pi}{2} -\frac{\sqrt{\pi\Theta }}{r} +\frac{(\pi\Theta )^{3/2}}{3r^{3}} +O(\Theta ^{5/2}), \\ 
\frac{r\sqrt{\pi\Theta }}{r^2 +\pi\Theta } &= \frac{\sqrt{\pi\Theta }}{r} -\frac{\pi^{3/2}\Theta ^{3/2}}{r^{3}} +O(\Theta ^{5/2}).
\end{align}

Substituting these expansions into equation \eqref{metric_exact} and retaining terms up to linear order in $\Theta $, the metric function reduces to
\begin{equation}
f(r)=1-\frac{\Lambda}{3}r^2-\frac{c}{r}+\frac{Q^2}{r^2}+\frac{8M}{\sqrt{\pi}}\frac{\sqrt{\Theta }}{r^2}-\frac{4Q^2}{\sqrt{\pi}}\frac{\sqrt{\Theta }}{r^3}+\mathcal{O}(\Theta ).
\end{equation}

In the commutative limit $\Theta  \to 0$, the metric function should reduce to the standard Reissner-Nordstrom-(A)dS form,
\begin{equation}
f(r) \xrightarrow{\Theta  \to 0}
1 - \frac{\Lambda}{3} r^2 - \frac{2 M}{r} + \frac{Q^2}{r^2} .
\end{equation}
Therefore we determine the constant
\begin{equation}
c = 2M.
\end{equation}
So that
\begin{equation}
f(r)=1-\frac{\Lambda}{3}r^2-\frac{2M}{r}+\frac{Q^2}{r^2}+\frac{8M}{\sqrt{\pi}}\frac{\sqrt{\Theta }}{r^2}-\frac{4Q^2}{\sqrt{\pi}}\frac{\sqrt{\Theta }}{r^3}
\label{metric_small_theta}
\end{equation}

Interestingly, even in the neutral case $Q=0$, a nonzero noncommutative parameter $\Theta$ generates an effective $1/r^{2}$ contribution to the metric function. This term plays a role similar to the electric charge term in the Reissner-Nordström geometry and can lead to the formation of an inner horizon. The horizons are determined by the condition $f(r)=0$. For $Q=0$ and neglecting the cosmological term, equation \eqref{metric_small_theta} reduces to
\begin{equation}
f(r)=r^{2}-2Mr+\frac{8M}{\sqrt{\pi}}\sqrt{\Theta}=0,
\end{equation}
and the spacetime therefore admits two distinct roots,
\begin{equation}
r_{\pm}=M\pm\sqrt{M^{2}-8M\sqrt{\frac{\Theta}{\pi}}},
\end{equation}
corresponding to the outer ($r_{+}$) and inner ($r_{-}$) horizons. 
In the regime where the noncommutative parameter is small, 
$\sqrt{\Theta}/M \ll 1$, the horizon radii can be expanded as
\begin{align}
r_+ &\approx 2M - 4 \sqrt{\frac{\Theta}{\pi}},\\
r_- &\approx 4 \sqrt{\frac{\Theta}{\pi}} .
\end{align}

The method employed in the following calculation was originally proposed in \cite{Urano_2009}. The mass parameter $M$ and the cosmological constant $\Lambda$ are fixed by simultaneously imposing the horizon conditions $f(r_+)=0$ and $f(r_c)=0$. To simplify the analysis, we introduce the dimensionless ratio
\begin{equation}
x \equiv \frac{r_+}{r_c}, \qquad 0 < x < 1.
\end{equation}

 In terms of $(r_c,x,Q,\Theta)$, the cosmological constant is given by 
\begin{multline}
\Lambda(r_c,x,Q,\Theta) = \frac{3(r_c^2 x - Q^2)}{r_c^4 x (1+x+x^2)} \\- \frac{12(1+x)(r_c^2 x - Q^2)} {\sqrt{\pi} r_c^5 x (1+x+x^2)^2} \sqrt{\Theta}.
\label{eq:Lambda}
\end{multline}

The corresponding mass parameter reads
\begin{multline}
M(r_c,x,Q,\Theta) = \frac{(1+x)\left[Q^2 + (Q^2 + r_c^2)x^2\right]} {2 r_c x (1+x+x^2)} \\
+ \frac{2\left[r_c^2(1+2x+2x^2+2x^3+x^4)+Q^2 x\right]}
{\sqrt{\pi} r_c^2 (1+x+x^2)^2} \sqrt{\Theta}.
\label{eq:Mass}
\end{multline}

The temperature associated with horizon is obtained from the surface gravity according to
\begin{equation}
T_+ = \frac{1}{4\pi} f'(r_+). 
\label{Temp1}
\end{equation}

\begin{equation}
T_c = -\frac{1}{4\pi} f'(r_c) 
\label{Temp2}
\end{equation}

Substituting \eqref{eq:Lambda} and \eqref{eq:Mass} into \eqref{Temp1}, \eqref{Temp2}  and rewriting the result in terms of $(r_c,x,Q,\Theta)$, we obtain the temperatures associated with the event horizon $T_+$ and the cosmological horizon $T_c$. The temperature of the event and cosmological horizons are respectively,
\begin{widetext}
\begin{multline}
T_+ = \frac{(x-1)\left[Q^2(1+2x+3x^2)-r_c^2 x^2(1+2x)\right]} {4\pi r_c^3 x^3(1+x+x^2)} \\ + \frac{(x-1)\left[r_c^2 x^2(1+x)(1+2x+3x^2) - Q^2(1+3x+6x^2+7x^3+4x^4)\right]} {\pi^{3/2} r_c^4 x^4 (1+x+x^2)^2} \sqrt{\Theta}.
\label{Tempplus}
\end{multline}
\begin{multline}
T_c = -\frac{(x-1)\left[r_c^2 x(2+x)-Q^2(3+2x+x^2)\right]} {4\pi r_c^3 x (1+x+x^2)} \\ - \frac{(x-1)\left[-r_c^2 x(1+x)(3+2x+x^2) + Q^2(4+7x+6x^2+3x^3+x^4)\right]} {\pi^{3/2} r_c^4 x (1+x+x^2)^2} \sqrt{\Theta}. \label{Tempc}
\end{multline}
\end{widetext}

In general, the temperatures associated with the event horizon and the cosmological horizon of a Reissner-Nordstrom-de Sitter  black hole are different. Consequently, the spacetime cannot, in general, be regarded as a thermodynamic equilibrium system.

Nevertheless, there exist special configurations in which both horizons share the same temperature. These include the Nariai black hole, for which the event and cosmological horizons coincide and the lukewarm black hole, in which the two distinct horizons possess equal and nonvanishing temperatures \cite{Romans:1991nq,Kastor:1993mj}. In such cases, a consistent thermodynamic description of the RN-dS spacetime becomes possible.

The lukewarm configuration corresponds to a particular equilibrium state in which the event horizon and the cosmological horizon have the same temperature. This condition is imposed by requiring
\begin{equation}
T_c = T_+ .
\label{Tempcondition}
\end{equation}

Enforcing this equality guarantees that no net heat flow occurs between the two horizons, allowing the RN-dS system to be treated as a genuine equilibrium thermodynamic configuration. Substituting the expressions for the cosmological constant \eqref{eq:Lambda} and the mass parameter \eqref{eq:Mass} into the temperature equality condition \eqref{Tempcondition}, we obtain the value of the electric charge required for the lukewarm state,
\begin{equation}
Q_{\mathrm{lw}}^{2} = \frac{r_c^{2} x^{2}}{(1+x)^{2}} - \frac{4 r_c x^{2} \left(2+3x+2x^{2}\right)} {\sqrt{\pi}(1+x)^{5}} \sqrt{\Theta}.
\label{Qlw}
\end{equation}

We observe that the electric charge required for the lukewarm configuration explicitly depends on the noncommutative parameter $\Theta$, indicating that noncommutative effects modify the thermal equilibrium condition of the RN-dS black hole. Once the lukewarm condition is satisfied, the two horizons share a common equilibrium temperature given by
\begin{equation}
T_{\mathrm{lw}} = \frac{1-x}{2\pi r_c (1+x)^2} + \frac{(x-1)(3+4x+3x^2)} {\pi^{3/2} r_c^{2} (1+x)^5} \sqrt{\Theta}.
\label{Tlw}
\end{equation}

In the commutative limit $\Theta \to 0$, together with $x \to 0$, the equilibrium temperature correctly reduces to the well-known de Sitter result $T = 1/(2\pi r_c)$. In this limit, the corresponding entropy also recovers the standard expression for de Sitter space \cite{GibbonsHawking1977,Sekiwa2006}.

\subsection{Physical region}

Not all of the black hole parameters result in a spacetime with black hole horizon(s). In general, the horizon structure of the spacetime is obtained from the roots of $f(r)=0$. For physically relevant configurations one expects three horizons, namely the inner (Cauchy) horizon $r_-$, the event horizon $r_+$, and the cosmological horizon $r_c$, satisfying $r_-<r_+<r_c$. However, since the metric function \eqref{metric_small_theta} is a quintic polynomial in $r$, the number and nature of its real roots depend sensitively on the parameters $(Q,\Theta)$. In particular, for some parameter values the equation may fail to produce distinct Cauchy and event horizons. It is therefore useful to identify the extremal configuration, where the Cauchy and event horizons coincide. This configuration marks the boundary of the physical region in parameter space where well-defined black hole solutions start to exist.

The minimum size of a black hole occurs when the black hole is extremal. This corresponds to the minimum value of $x$, below which there is only cosmological horizon. The extremal configuration is obtained by imposing the conditions $f(r)=0$ and $f'(r)=0$ on the metric function \eqref{metric_small_theta}. The parameters $M$ and $\Lambda$ appearing in the metric are determined from equations \eqref{eq:Mass} and \eqref{eq:Lambda}, which follow from the horizon conditions $f(r_+)=0$ and $f(r_c)=0$. We note that the extremality conditions determine $x_{\rm min}$~(minimum of $x$) as a function of the charge $Q$ and the noncommutative parameter $\Theta$.

\begin{figure}[H]
    \centering
    \includegraphics[width=1\linewidth]{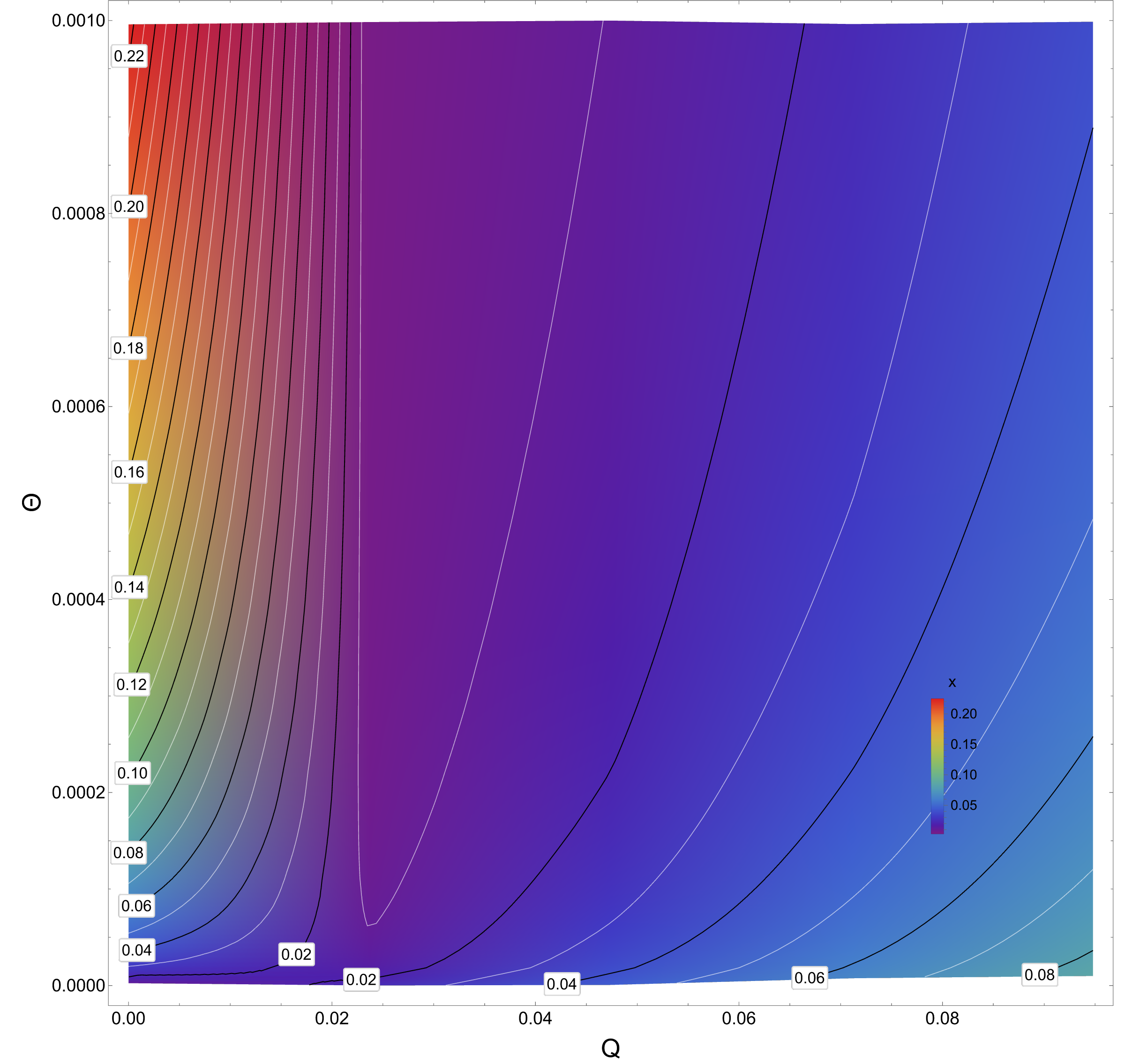}
    \caption{Physical region in the $(Q,\Theta)$ plane with $r_c=1$. The colour scale represents the numerical extremal horizon ratio $x_{\rm min}$, above which black hole horizon exists. The black and white curves represent the contour lines of constant $x_{\rm min}$.}  
    \label{cont}
\end{figure}

Since the extremality equations cannot be solved analytically, we determine the extremal values of $x$ numerically for $0\le Q\le0.1$ and $0\le\Theta\le10^{-3}$. The resulting parameter space is shown in Fig.~\ref{cont}, where the colour scale represents the numerically obtained values of $x_{\rm min}$ and the black curves denote contours of constant $x_{\rm min}$. In the neutral limit $Q=0$, the extremal configuration is governed entirely by the noncommutative parameter $\Theta$. As $\Theta$ increases, the extremal horizon ratio $x$ also increases. This behaviour can be understood from the additional positive correction $\sim \sqrt{\Theta}/r^2$ appearing in the metric function \eqref{metric_small_theta}, which shifts the extremal solution toward larger values of the black hole horizon $r_+$ relative to the cosmological horizon $r_c$, thereby increasing the ratio $x=r_+/r_c$. The black contour lines correspond to $x_{\rm min}= 0.02, 0.04, 0.06, \dots$ whereas the white contour lines correspond to $x_{\rm min}=0.01, 0.03, 0.05, \dots$

\subsection{Entropy}
To calculate entropy {\it at a fixed} $\Lambda$, we have to express mass and temperature as functions of $r_{+}$ and $\Lambda$. Imposing the condition $f(r_+)=0$, we obtain the mass function
\begin{equation}
M(r_+, Q, \Lambda, \Theta) = \frac{3r_+\left(Q^{2} + r_+^{2}\right) - r_+^{5}\Lambda}{6r_+^{2}}+\frac{2\sqrt{\Theta}\left(3 - r_+^{2}\Lambda\right)}{3\sqrt{\pi}},
\label{Ms}
\end{equation}
and the temperature, via $T_+=\frac{1}{4\pi}f'(r) \Big{|}_{r_+}$, 
\begin{multline}
T_+(r_+,Q,\Lambda,\Theta) = \frac{(3Q^2 - 4Mr_+)\sqrt{\Theta}}{\pi^{3/2} r_+^4}\\ - \frac{3Q^2 - 3Mr_+ + r_+^4\Lambda}{6\pi r_+^3}.
\label{Ts}
\end{multline}

The entropy associated with each horizon can be obtained from the first law of black hole thermodynamics. At fixed electric charge $Q$ and cosmological constant $\Lambda$, the entropy is defined as
\begin{equation}
S_+ = \int \frac{1}{T_+} \left( \frac{\partial M}{\partial r_+} \right)_{Q,\Lambda} dr_+.
\label{eq:Sh_def}
\end{equation}

To be clear, we note that $M(r_+)$ and $T_+(r_+)$ in the expression \eqref{eq:Sh_def} are defined in equation \eqref{Ms} and \eqref{Ts}, after which we expand in the small-$\Theta$ regime and retain the terms up to $\mathcal{O}(\Theta)$. The entropy takes the form,
\begin{align}
S_+
&= \int_{0}^{r_+} \left[ 2\pi r + 8\sqrt{\pi\Theta} \right] dr \nonumber\\
&= \pi r_+^2 + 8\sqrt{\pi\Theta}r_+ \nonumber\\
&= x^2 \pi r_c^2 + 8x r_c \sqrt{\pi\Theta},  \label{S0}
\end{align}
as obtained in \cite{Hamil_2025}. 
Similarly, to compute the cosmological entropy at fixed cosmological constant $\Lambda$, 
we express both the mass and the temperature as functions of 
$r_c$ and $\Lambda$. Imposing the horizon condition $f(r_c)=0$, 
the mass parameter can be written as
\begin{equation}
M(r_c, Q, \Lambda, \Theta) = \frac{3r_c\left(Q^{2} + r_c^{2}\right) - r_c^{5}\Lambda}{6r_c^{2}}+\frac{2\sqrt{\Theta}\left(3 - r_c^{2}\Lambda\right)}{3\sqrt{\pi}},
\end{equation}
and the temperature, via $T_c=-\displaystyle{\frac{1}{4\pi}}f'(r) \Big{|}_{r_c}$, 
\begin{equation}
T_c(r_c,Q,\Lambda,\Theta) = \frac{(3Q^2 - 4Mr_c)\sqrt{\Theta}}{\pi^{3/2} r_c^4} - \frac{3Q^2 - 3Mr_c + r_c^4\Lambda}{6\pi r_c^3}.
\label{Tc}
\end{equation}
The entropy reads
\begin{align}
S_c = \int \frac{1}{T_c} \left( \frac{\partial M}{\partial r_c} \right)_{Q,\Lambda} dr_c, 
\end{align}
So that
\begin{align}
S_c
&= \int_{0}^{r_c} \left[ 2\pi r + 8\sqrt{\pi\Theta} \right] dr \nonumber\\
&= \pi r_c^2 + 8\sqrt{\pi\Theta}r_c 
\end{align}
Another approach to compute the entropy of the black hole is to allow variation in $\Lambda$ and integrate entropy using mass, Eqn.\eqref{eq:Mass}, and $T_{+}$, Eqn.\eqref{Tempplus}, according to
\begin{eqnarray}
S &=&\int^{r_{+}}_{0} \frac{1}{T_{+}(x)} \left( \frac{\partial M}{\partial r_{+}} \right)_{Q} d r_{+}=\int^{x}_{0} \frac{1}{T_{+}(x)}\left( \frac{\partial M(x)}{\partial x} \right)_{Q} dx,  \nonumber\\ 
&=& \pi r_{c}^{2}\int^{x}_{0}\frac{2x}{x^{3}-1} dx - r_{c}\sqrt{\Theta}\int^{x}_{0}\frac{8 \sqrt{\pi }\left(x^3+x^2+x+1\right)}{(x-1) \left(x^2+x+1\right)^2}  dx, \nonumber
\end{eqnarray}
resulting in
\begin{equation}
S(x) = \pi r_{c}^2 \mathcal{F}_{1}(x)+8r_{c}\sqrt{\pi\Theta}\mathcal{F}_{2}(x),  \label{Sformula}   
\end{equation}
where 
\begin{eqnarray}
    \mathcal{F}_{1}(x)=&&\frac{\pi}{3\sqrt{3}}+\frac{1}{3} \log \left(x^2+x+1\right)-\frac{2}{3} \log (1-x) \notag \\
    &&-\frac{2}{\sqrt{3}}\tan ^{-1}\left(\frac{2 x+1}{\sqrt{3}}\right), \nonumber\\
    \mathcal{F}_{2}(x)=&&\frac{3 x}{9(x^2+x+1)}+\frac{2}{9} \log \left(\frac{x^2+x+1}{x^2-2x+1}\right).  \nonumber
\end{eqnarray} 
Remarkably, for small $x$ up to the leading order $\mathcal{O}(x^{2})$,
\begin{equation}
    S(x) \simeq \pi r_{c}^{2}x^{2}+8x r_{c}\sqrt{\pi\Theta},  \nonumber
\end{equation}
exactly the entropy given by Eqn.(\ref{S0}). In the commutative limit $\Theta \to 0$, this expression correctly reduces to the Bekenstein-Hawking area law $S = \pi r_+^2$ \cite{Mann:2025xrb}. 

For the entropy of cosmological horizon, we start with the relation~\cite{Urano_2009}
\begin{eqnarray}
    S_{c} =&&- \int_{r_{+}}^{r_{c}}\left(\frac{dM}{T_{c}}+\frac{1}{T_{c}}\frac{r_{c}^{3}}{6}\frac{\partial\Lambda}{\partial r_{c}}\right)~dr_{c}, \notag \\
      =&& -\int_{1}^{x}\frac{1}{T_{c}(x)}\left( \frac{\partial M(x)}{\partial x}+\frac{r_{+}^{3}}{6x^{3}}\frac{\partial \Lambda(x)}{\partial x}\right)~dx, \notag \\  \label{Scintegral}
\end{eqnarray}
where $M(x), \Lambda(x), T(x)$ are given by substituting $r_{c}=r_{+}/x$ in Eqn.~(\ref{eq:Mass}),(\ref{eq:Lambda}) and (\ref{Tempc}) respectively, i.e., the integration is performed at a fixed black hole horizon $r_{+}$. We then obtain
\begin{eqnarray}
    S_{c}(x)&&= \frac{\pi r_{+}^{2}(1-x^2)}{x^2} + \frac{4 r_{+}}{3}\sqrt{\pi\Theta } \Bigg[\frac{2\pi}{\sqrt{3}}-2 \log (1-x)\Big{|}_{1^{+}}^{x} \notag\\
    &&-2\sqrt{3}  \tan ^{-1}\left(\frac{2 x+1}{\sqrt{3}}\right)+\log \left(\frac{x^2+x+1}{3}\right)\Bigg]. \label{Scformula}
\end{eqnarray}
Interestingly, the noncommutativity induces a divergent term $\sim \log(0^{+})$ in the Nariai limit since $T_{c}$ becomes zero, while the divergent terms from zero temperature cancel one another for the GR terms. In the small black hole and no noncommutativity limit,~$x\to 0,\theta \to 0$, Eqn.~(\ref{Scformula}) gives
\begin{equation}
    S_{c}(0)=\pi r_{c}^{2},
\end{equation}
the Gibbons-Hawking area law of the cosmological-horizon entropy. 

The total entropy $S(x)+S_{c}(x)$ from Eqn.~(\ref{Sformula}) and (\ref{Scformula}) still contains divergence in the Nariai limit $x\to 1$ due to the effect of zero temperature. As we will see in the following, this divergence of entropy, originated from the entanglement between two horizons remain even in the effective thermodynamic picture. 

\section{Effective Thermodynamics}
In a spacetime containing both a black hole event horizon and a cosmological horizon, thermodynamics cannot be characterized by a single local horizon quantity. Instead, the system must be regarded as a composite configuration consisting of two interacting horizons. A meaningful thermodynamic description can therefore be formulated in terms of effective, global quantities that capture the collective behavior of the system as a whole.

We adopt an effective first law of thermodynamics of the form \cite{Urano_2009,Nakarachinda_2021}
\begin{equation}
dM = T_{\mathrm{eff}}dS - P_{\mathrm{eff}}dV + \phi_{\mathrm{eff}}dQ ,
\label{firstlaw}
\end{equation}
where the mass parameter $M$ is interpreted as the internal energy of the combined two-horizon system. The thermodynamic volume is naturally identified with the geometric volume enclosed between the event horizon $r_+$ and the cosmological horizon $r_c$, namely
\begin{equation}
V = \frac{4\pi}{3} r_c^{3}\left(1 - x^3 \right).
\label{volume}
\end{equation}
Since the physical system occupies the region between the two horizons, the allowed range of the parameter $x$ is $0 < x < 1$.

In conventional treatments of RN-dS thermodynamics, the total entropy is assumed to be the sum of the individual horizon entropies,
\begin{equation}
    S_{\mathrm{total}} = S_{+} + S_{c}.
\end{equation}
Under this assumption, the effective temperature takes the form, $T_{\mathrm{eff}} = \frac{T_{+} T_{c}}{T_{+} + T_{c}} $ \cite{Urano_2009}. However, this expression becomes problematic in the lukewarm limit, where $T_{+} = T_{c} = T$. In this case, it yields $T_{\mathrm{eff}}^{\mathrm{lw}} = T^{\mathrm{lw}}/2$, which contradicts the expectation for a system in genuine thermal equilibrium.

To resolve this inconsistency, we introduce an additional correlation contribution to the entropy that encodes the interaction between the event horizon and the cosmological horizon. Including the leading noncommutative correction, we write the total entropy as, which is motivated by \cite{Zhang_2016,Ma2022}
\begin{align}
S(x) &= S_{c} + S_{+} + S_{\mathrm{int}} \nonumber\\
&\equiv \pi r_{c}^{2} f_{1}(x)
+ 8 r_{c}\sqrt{\pi\Theta}f_{2}(x),
\label{entropy}
\end{align}
where the functions $f_{1}(x)$ and $f_{2}(x)$ characterize the correlation effects between the two horizons.

The lukewarm configuration provides a natural and physically well-motivated criterion for fixing the correlation functions appearing in the total entropy. In this configuration, the black hole and cosmological horizons share a common temperature. We therefore require that the effective temperature derived from the entropy \eqref{entropy} reproduces this common horizon temperature in the
lukewarm limit,
\begin{equation}
T_{\mathrm{eff}}^{\mathrm{lw}} = T_{\mathrm{lw}}  .
\label{lukewarm}
\end{equation}

For a thermodynamic system parametrized by the variables $(x,r_c)$, the effective temperature follows from the first law and can be expressed in Jacobian form as
\begin{equation}
T_{\mathrm{eff}}(x,r_c) = \left(\frac{\partial M}{\partial S}\right)_{V,Q} =\left(\frac{ \partial_x M \partial_{r_c} V - \partial_{r_c} M \partial_x V}{ \partial_x S \partial_{r_c} V - \partial_{r_c} S \partial_x V}\right)_{V,Q} \label{Teqn}
\end{equation}

Substituting the explicit expressions for the mass, entropy and thermodynamic volume, the lukewarm effective temperature takes the form
\begin{widetext}
    \begin{multline} 
T_{\mathrm{eff}}^{\mathrm{lw}} = \frac{1+x^{4}}{A(x)} \Bigg\{ -\sqrt{\pi}r_cx (1+x)^{3} (1+x+x^{2}) \Bigl[(x^{3}-1) f_{1}'(x) - 2x^{2} f_{1}(x)\Bigr] \\ - 2\sqrt{\Theta} \Bigg[ -2x^{2}\,\mathcal{P}(x)f_{1}(x) + (1+x+x^{2}) \Bigl( 4x^{3}(1+x)^{3} f_{2}(x) \\ \qquad\qquad + (x-1)\mathcal{P}(x) f_{1}'(x) - 4x(1+x)^{3}(x^{3}-1) f_{2}'(x) \Bigr) \Bigg] \Bigg\}, 
\end{multline}
\end{widetext}
with
\begin{equation}
\mathcal{P}(x)=2 + x(1+x)\bigl(5 + x(2 + x(4 + x(3+2x)))\bigr),
\end{equation}
and
\begin{equation}
A(x) = \pi^{3/2} r_c^{2} (1+x)^{5} (1+x+x^{2})^{2} \Bigl[ (x^{3}-1) f_{1}'(x) - 2x^{2} f_{1}(x) \Bigr]^{2}.
\end{equation}

In the limit where only the cosmological horizon remains ($x \to 0$), correlation
effects between the two horizons must vanish. Requiring the entropy to reduce to
that of an isolated cosmological horizon imposes the boundary conditions
\begin{equation}
f_{1}(0)=1,
\qquad
f_{2}(0)=1 .
\label{condition}
\end{equation}
With these conditions, the total entropy satisfies
\begin{equation}
S(x=0)=\pi r_c^{2}+8 r_c \sqrt{\pi\Theta},
\end{equation}
which coincides with the cosmological horizon entropy.

The functions $f_1(x)$ and $f_2(x)$ are uniquely determined in the physical domain $0<x<1$ as,
\begin{align}
f_{1}(x) &= \frac{3 - 5x^{2} - 5x^{3} + 3x^{5}}{5(x^{3}-1)} + \frac{8}{5}(x^{3}-1)^{2/3}, \\ 
f_{2}(x) &= -\frac{9 - 5x - 5x^{2} - 26x^{3} - 5x^{4} - 5x^{5} + 9x^{6}} {14(1-x)(1+x+x^{2})^{2}} \nonumber\\&\phantom{=}- \frac{23}{14}(x^{3}-1)^{1/3}.
\label{solutionf}
\end{align}
We note that the function $f_{1}(x)$ is the result reported in \cite{Zhang_2016}.

With the correlation functions fixed, the lukewarm effective temperature reduces to the compact form
\begin{equation}
T_{\mathrm{eff}}^{\mathrm{lw}}(r_c,x,\Theta) = \frac{1-x}{2\pi r_c (1+x)^{2}} \left[ 1 - \frac{2(3+4x+3x^{2})}{\sqrt{\pi} r_c (1+x)^{3}} \sqrt{\Theta} \right].
\label{Teff}
\end{equation}

Finally, we summarize our main results. Once the functions $f_{1}(x)$ and
$f_{2}(x)$ have been fixed by imposing the lukewarm condition, we are able
to write down explicit expressions for the effective thermodynamic
quantities of the system. In particular, we obtain closed-form expressions
for the effective temperature, effective pressure and effective electric
potential.

The effective temperature is given by
\begin{multline}
T_{\mathrm{eff}} = -\frac{(x-1)}{\mathcal{G}(x)} \Bigg[ r_c^{2} x^{2} \Big( \sqrt{\pi}r_c x (1+2x+2x^{5}+x^{6}) \\ -4(1+2x+2x^{2}-x^{3}-x^{4}+2x^{5}+2x^{6}+x^{7}) \sqrt{\Theta} \Big) \\ +Q^{2} \Big( -\sqrt{\pi}r_c x (1+2x+3x^{2}+3x^{6}+2x^{7}+x^{8}) \\ +4(1+2x+3x^{2}+2x^{3}-x^{4}-x^{5}+2x^{6}+3x^{7}+2x^{8}+x^{9}) \sqrt{\Theta} \Big) \Bigg],
\end{multline}
where the function $\mathcal{G}(x)$ is given as follows
\begin{equation}
\mathcal{G}(x)\equiv 4 \pi^{3/2} r_c^{4} x^{4} (1+x+x^{2})(1+x^{4}).
\end{equation}
The corresponding effective pressure takes the form
\begin{multline}
P(r_c,x,Q,\Theta)=\mathcal{K}(x)
\Big[
r_c^{2}x^{2}\,A(x;u)+Q^{2}\,B(x;u) \\
+2\sqrt{\Theta}\Big(-r_c^{2}x^{2}\,C(x;u)+Q^{2}\,D(x;u)\Big)
\Big]
\end{multline}
The full expression for the effective pressure is presented in Appendix \ref{app:P_full}.

Finally, the effective electric potential is found to be
\begin{equation}
\phi_{\text{eff}} = \frac{Q\left[ r_c (1+x)(1+x^{2})(1+x+x^{2}) + \frac{4 x^{2}\sqrt{\Theta}}{\sqrt{\pi}} \right]}{r_c^{2} x (1+x+x^{2})^{2}} .
\end{equation}

\begin{figure}[h]
    \centering
    \includegraphics[width=0.9\linewidth]{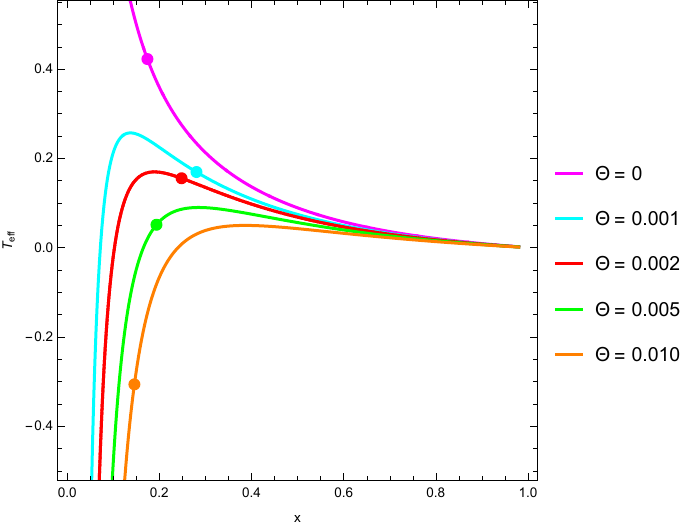}
    \caption{Behavior of the effective temperature $T_{\mathrm{eff}}(x)$ for the Schwarzschild--de Sitter black hole ($Q=0$) with $r_c=1$, shown for several values of the noncommutative parameter $\Theta$. For each curve, the marked point represents the minimum position $x_{\min}$.}
    \label{fig:zerocharge}
\end{figure}

In the simplest case illustrated in Fig.~\ref{fig:zerocharge}, corresponding to the Schwarzschild-dS black hole, the effective temperature never reaches zero in the commutative limit ($\Theta=0$) except in the Nariai limit where the black hole horizon overlaps with cosmological horizon. The presence of even a tiny non-commutative parameter $\Theta$ inevitably drives the effective temperature into the negative regime, but these are not physical since $x<x_{\rm min}$ established in Fig.~\ref{cont}. There is no actual black hole horizon for these values of $x$ that give negative effective temperature.  

\begin{figure}[H]
    \centering
    \includegraphics[width=0.9\linewidth]{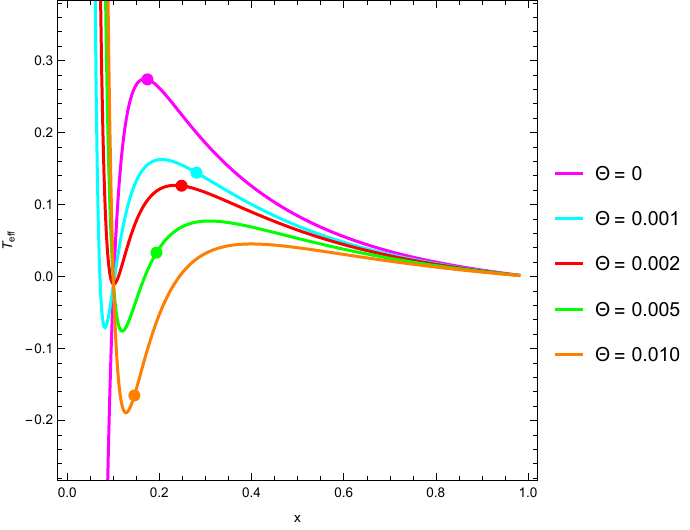}
    \caption{Behavior of the effective temperature $T_{\mathrm{eff}}$ 
    as a function of $x$ for $Q = 0.1$, $r_c=1$, and different values 
    of the noncommutative parameter $\Theta$. For each curve, the marked point represents the minimum position $x_{\min}$.}
    \label{fig:Teffvaryth}
\end{figure}

We now examine the consistency of our results in Fig.~\ref{fig:Teffvaryth} with the analysis of Ref.~\cite{Zhang_2016}. Although the same entropy modification scheme is employed, our study extends the discussion to the noncommutative case $\Theta \neq 0$. In the commutative limit $\Theta = 0$, represented by the pink curve, our result reproduces the behavior reported in Ref.~\cite{Zhang_2016}, thereby providing a consistency check of the formalism. Let us first consider the commutative case. We observe that $T_{\mathrm{eff}}$ can become negative in certain regions of $x$. However, there is actually no black hole for these spacetime parameters. The value of $x$ is below $x_{\rm min}$ given in Fig.~\ref{cont}, outside the physical region where there is black hole horizon.

In the lukewarm configuration, where the event and cosmological horizons are in thermal equilibrium, one obtains
\[ T_{\mathrm{eff}} = T_c = T_+ , \] 
as expected for two subsystems sharing a common temperature.

This contrasts with the extended Iyer-Wald formalism \cite{Urano_2009}, where the total entropy is taken as $S = S_c + S_+$ without additional correlation terms. In that framework, the effective temperature remains strictly positive but reduces to
\begin{equation}
T_{\mathrm{eff}} = \frac{T_+}{2} = \frac{T_c}{2}.
\end{equation}

Such a prescription does not fully capture the physical meaning of thermal equilibrium: if two systems are each at $50\,\mathrm{K}$, their combined equilibrium temperature should remain $50\,\mathrm{K}$, not $25\,\mathrm{K}$. The correlation-modified entropy therefore provides a more consistent thermodynamic interpretation. Although $T_{\mathrm{eff}}$ becomes negative in certain range of $x$, it is found that all the negative temperature occurs when $x<x_{\rm min}$~(given in Fig.~\ref{cont}). Thus, there is actually no black hole horizon and no thermodynamics for this case.

Figure~\ref{fig:Teffvaryth} shows the effective temperature $T_{\mathrm{eff}}$ as a function of the dimensionless parameter $x$ for fixed $Q=0.1$ and several values of the noncommutative parameter $\Theta$. In the commutative limit ($\Theta=0$), the temperature is negative at very small $x$, then rises rapidly, becomes positive, reaches a local maximum, and gradually decreases toward zero as $x \to 1$. Again, the negative effective temperature region is unphysical since $x<x_{\rm min}$ and there is no black hole being formed. 

Once noncommutative effects are included, the behavior changes in a noticeable way. The positive peak is progressively reduced, and a deeper negative minimum develops in the small-to-intermediate range of $x$. As $\Theta$ increases, this negative branch becomes more pronounced and extends over a wider interval, indicating that noncommutativity strongly reduces the physical region $x\geq x_{\rm min}$.

As $x\to 1$, however, all curves converge smoothly toward zero, showing the asymptotic Nariai behavior. The impact of the noncommutative parameter is therefore most significant away from the asymptotic limit, where it substantially reshapes the thermal structure of the system.

\section{Thermodynamics Stability}
In this section, we investigate the thermodynamic stability of the system by examining the behavior of the entropy, heat capacity and Helmholtz free energy. The analysis is performed in the canonical ensemble, where the electric charge $Q$ and the thermodynamic volume $V$ are held fixed.

Among these quantities, the heat capacity plays a central role, as it directly controls the response of the black hole to thermal fluctuations. A positive heat capacity indicates a locally stable equilibrium, in which the absorption of energy leads to an increase in temperature, consistent with ordinary thermodynamic systems. In contrast, a negative heat capacity reflects an intrinsically unstable regime, where energy absorption results in cooling, signaling a runaway behavior characteristic of self-gravitating systems \cite{Mann:2025xrb,Singh:2025svv}. The Helmholtz free energy provides complementary information about global stability, allowing different thermodynamic branches to be compared and possible phase transitions to be identified. Together with the entropy behavior, these quantities furnish a complete picture of both local and global thermodynamic stability.


\subsection{Heat Capacity}
Now, let us begin by deriving the heat capacity at constant volume, $C_V$, which provides a direct diagnostic of local thermodynamic stability in the canonical ensemble. By definition, it is given by
\begin{equation}
C_V = T_{\mathrm{eff}} \left(\frac{\partial S}{\partial T_{\mathrm{eff}}}\right)_{V,Q}.
\end{equation}

To implement the constant-volume constraint explicitly, we employ the Jacobian method and reexpress $C_V$ in terms of derivatives with respect to the variables $(r_c,x)$. This yields
\begin{equation}
C_V = \left( \frac{ \partial_x M\,\partial_{r_c} V - \partial_{r_c} M\,\partial_x V }{ \partial_x S\,\partial_{r_c} V - \partial_{r_c} S\,\partial_x V } \right)_{Q} \left( \frac{\partial_x S}{\partial_x T_{\mathrm{eff}}} \right)_{V,Q}.
\end{equation}

To isolate the effects of spacetime noncommutativity, we expand the heat capacity in powers of the noncommutative parameter $\Theta$, obtaining
\begin{widetext}
\begin{gather}
C_V(r_c,x,Q,\Theta)=C_0+C_{1/2}\sqrt{\Theta},
\end{gather}
where,
\begin{gather}
C_{0} = -\,\frac{ 2 \pi r_c^{2} x^{2} (1+x^{4})^{2} \Big[ r_c^{2} x^{2}\,\mathcal{N}_{1}(x) - Q^{2}\,\mathcal{N}_{2}(x) \Big] }{ (x^{2}-1)(1+x+x^{2}) \Big[ r_c^{2} x^{2}\,\mathcal{D}_{1}(x) - Q^{2}\,\mathcal{D}_{2}(x) \Big] },\\
C_{1/2} = \frac{ 16 \sqrt{\pi}r_c x (1+x^{4})^{2} \Big[ r_c^{4} x^{4}\,\mathcal{P}_{1}(x) - Q^{2} r_c^{2} x^{2}\,\mathcal{P}_{2}(x) + 2 Q^{4}\,\mathcal{P}_{3}(x) \Big] }{ (x-1)(1+x)^{2}(1+x+x^{2})^{2} \Big[ r_c^{2} x^{2}\,\mathcal{D}(x) - Q^{2}\,\mathcal{E}(x) \Big]^{2} }.
\end{gather}
\end{widetext}


Here, $\mathcal{N}_{i}(x)$, $\mathcal{D}_{i}(x)$, $\mathcal{P}_{i}(x)$ and $\mathcal{E}(x)$ denote polynomial functions of $x$ introduced to streamline
the presentation. Their explicit forms are collected in
Appendix \ref{app:Heat}.

In Fig. \ref{fig:HeatCapacity}, we plot the heat capacity $C_V$ as a function of the horizon ratio $x$ and the effective temperature $T_{\mathrm{eff}}$ for fixed charge $Q=0.3$ and thermodynamic volume $V=3$, and for $\Theta = 0,\,0.005,\,0.010,\,0.015$ and $0.020$. The dots indicate the lower bound of the physical region, corresponding to the extremal configuration as analyzed in the previous section. We note that the case with $\Theta$ has a numerical lower bound  at $x_{\text{min}}=-9.11\times10^{-9}\approx0$, therefore, the dot is not shown.

The heat capacity is separated into two distinct branches by an asymptote at which $C_V$ diverges, signaling the presence of a phase transition. The arrows indicate the direction of increasing $x$, which serves as an implicit thermodynamic parameter along the curve. note that the divergence occurs at relatively large values of $x$, well beyond the lower bound of the physical region determined by the extremal configuration, indicating that the phase transition takes place far from the boundary of the allowed parameter space.

Moreover, the small-$x$ black hole branch exhibits a region of positive heat capacity, indicating local thermodynamic stability. As the noncommutative parameter $\Theta$ increases, this stable region is progressively suppressed, with the magnitude of the heat capacity decreasing. In contrast, the large-$x$ black hole branch is characterized by negative heat capacity, corresponding to an unstable configuration and this instability is further enhanced as $\Theta$ increases. These results demonstrate that noncommutative effects significantly modify the stability structure and phase behavior of the system.

\begin{figure}[htbp]
\centering

\subfloat[$C_V$-$x$]{%
    \includegraphics[width=0.9\linewidth]{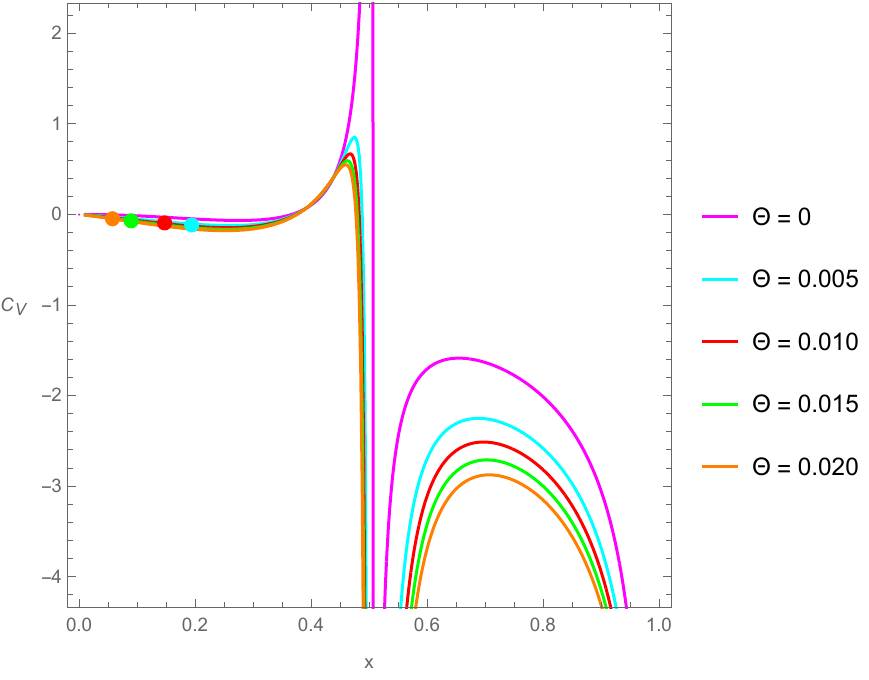}
}
\hfill
\subfloat[$C_V$-$T_{\mathrm{eff}}$]{%
    \includegraphics[width=0.9\linewidth]{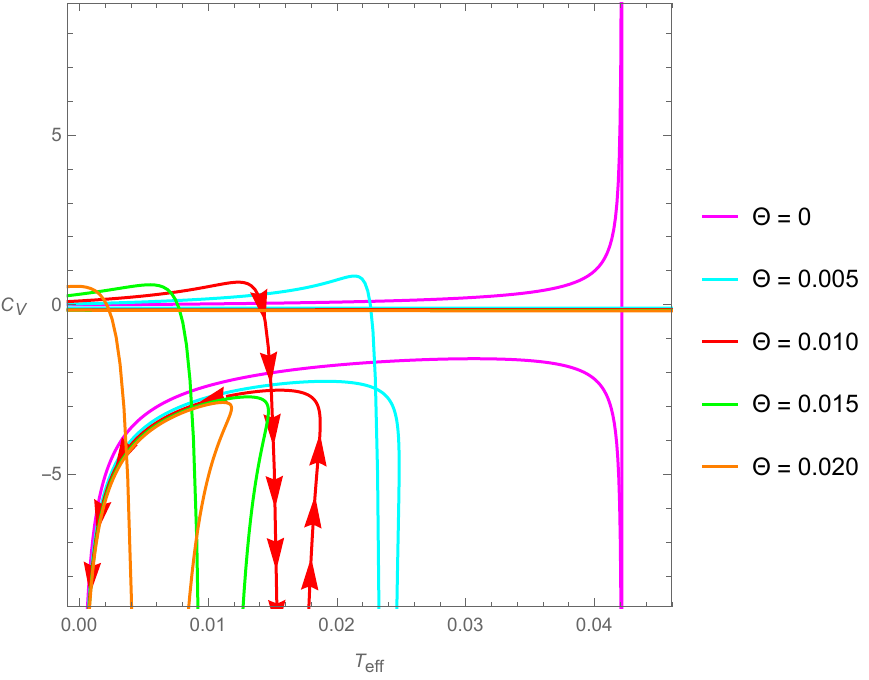}
}

\caption{Heat capacity $C_V$ as a function of $x$ and the effective temperature $T_{\mathrm{eff}}$ for fixed $Q=0.3$, $V=3$ with $\Theta = 0,\,0.005,\,0.010,\,0.015,$ and $0.020$. }
\label{fig:HeatCapacity}

\end{figure}

\subsection{Entropy}
\begin{figure}[htbp]
\centering

\subfloat[$S$-$x$]{%
    \includegraphics[width=0.9\linewidth]{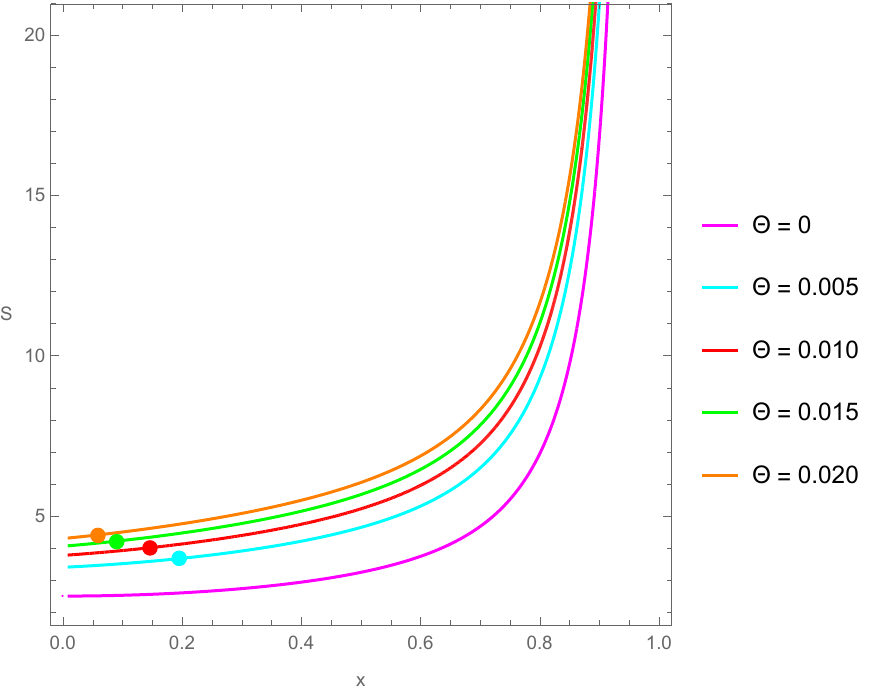}
}
\hfill
\subfloat[$S$-$T_{\mathrm{eff}}$]{%
    \includegraphics[width=0.9\linewidth]{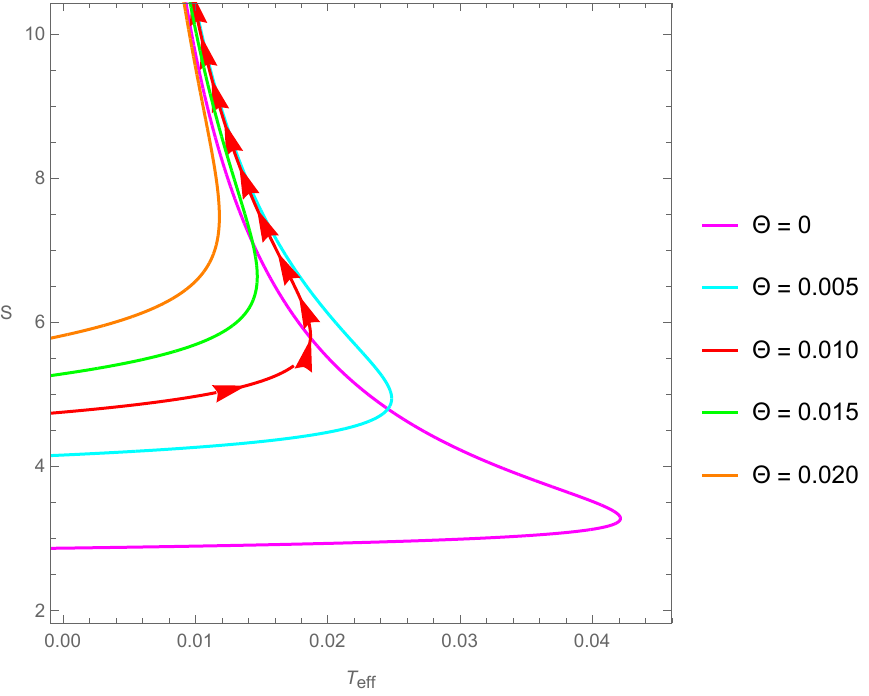}
}

\caption{
Entropy $S$ as a function of the effective temperature 
$T_{\mathrm{eff}}$ and the parameter $x$ for fixed charge 
$Q=0.3$, $V=3$ with $\Theta = 0,\,0.005,\,0.010,\,0.015,$ and $0.020$. }
\label{fig:Entropy}

\end{figure}

Fig.~\ref{fig:Entropy} shows the entropy $S$ as a function of the effective temperature $T_{\mathrm{eff}}$ for fixed charge $Q=0.3$ and thermodynamic volume $V=3$, and for the same values of the noncommutative parameter $\Theta = 0,\,0.005,\,0.010,\,0.015$ and $0.020$ used in the heat capacity analysis. Smaller black holes, corresponding to lower values of $x$, are associated with lower entropy and higher effective temperatures, whereas larger black holes occupy the opposite regime of higher entropy and lower temperature. Importantly, the entropy remains continuous over the entire temperature range, absence of any entropy discontinuity across the transition. The critical behavior is encoded in the curvature of the entropy profile, i.e. an inflection point appears at the critical temperature, coinciding with the divergence of the heat capacity. 

\subsection{Helmholtz Free Energy}
We begin by examining the response of the effective Helmholtz free energy $F$ to variations in the effective temperature $T_{\mathrm{eff}}$, treating the entropy $S$ as an implicit parameter. Applying the chain rule, one may write
\begin{equation}
\left(\frac{\partial F(S)}{\partial T_{\mathrm{eff}}(S)}\right)_{V,Q} = \left(\frac{\partial F}{\partial S}\right)_{V,Q} \left(\frac{\partial T_{\mathrm{eff}}}{\partial S}\right)_{V,Q}^{-1}.
\label{Fim}
\end{equation}

From the effective first law \eqref{firstlaw}, the effective temperature is
defined as
\begin{equation}
T_{\mathrm{eff}} = \left(\frac{\partial M}{\partial S}\right)_{V,Q},
\label{Teffdef}
\end{equation}
which immediately implies
\begin{equation}
\left(\frac{\partial T_{\mathrm{eff}}}{\partial S}\right)_{V,Q} = \left(\frac{\partial^2 M}{\partial S^2}\right)_{V,Q}.
\label{Teffder}
\end{equation}

Substituting equation \eqref{Teffder} into \eqref{Fim}, we obtain
\begin{equation}
\left(\frac{\partial F}{\partial T_{\mathrm{eff}}}\right)_{V,Q} = \left(\frac{\partial F}{\partial S}\right)_{V,Q} \left(\frac{\partial^2 M}{\partial S^2}\right)_{V,Q}^{-1}.
\label{predict1}
\end{equation}

The second derivative of the mass with respect to entropy can be expressed in terms of the heat capacity at fixed volume,
\begin{equation}
C_V = T_{\mathrm{eff}} \left(\frac{\partial S}{\partial T_{\mathrm{eff}}}\right)_{V,Q} = \frac{T_{\mathrm{eff}}} {\left(\frac{\partial^2 M}{\partial S^2}\right)_{V,Q}}.
\label{CVdef}
\end{equation}
Using this relation, equation \eqref{predict1} can be written in the compact form
\begin{equation}
\left(\frac{\partial F}{\partial T_{\mathrm{eff}}}\right)_{V,Q} = \left(\frac{\partial F}{\partial S}\right)_{V,Q} \frac{C_V}{T_{\mathrm{eff}}}.
\label{predict}
\end{equation}

We now turn to the critical behavior of the system. The critical point is identified by the conditions
\begin{gather}
\left(\frac{\partial T_{\mathrm{eff}}}{\partial S}\right)_{V,Q} = \left(\frac{\partial^2 T_{\mathrm{eff}}}{\partial S^2}\right)_{V,Q} = 0 ,
\label{criticalcond1}
\end{gather}
which signal a qualitative change in the thermodynamic response.

When $\partial T_{\mathrm{eff}}/\partial S = 0$, or equivalently when $\partial S/\partial T_{\mathrm{eff}}$ diverges, an infinitesimal change in temperature produces a large variation in entropy. Consequently, the heat capacity $C_V$ diverges, indicating the breakdown of thermal stability and the onset of critical behavior. Despite this divergence, the entropy itself remains continuous across the transition. Rather than exhibiting a discontinuity, the $S$-$T_{\mathrm{eff}}$ curve develops an inflection point at criticality.

Equation \eqref{predict} shows that the temperature derivative of the Helmholtz free energy is directly governed by the heat capacity. At the critical point where $C_V$ diverges, the slope $\left(\partial F/\partial T_{\mathrm{eff}}\right)_{V,Q}$ becomes singular. This signals a loss of analyticity in the first derivative of the free energy, even though the free energy itself remains continuous. According to Ehrenfest's classification, the absence of any discontinuity in the entropy excludes a first-order phase transition. Instead, the singular behavior of $\partial F/\partial T_{\mathrm{eff}}$, together with the divergence of $C_V$ and the existence of inflection point in the $S$-$T_{\mathrm{eff}}$ curve, identifies the transition as second order. Consequently, the $F$-$T_{\mathrm{eff}}$ diagram is expected to develop a cusp or kink at the critical temperature.

\begin{figure}[htbp]
\centering

\subfloat[$F$-$x$]{%
    \includegraphics[width=0.9\linewidth]{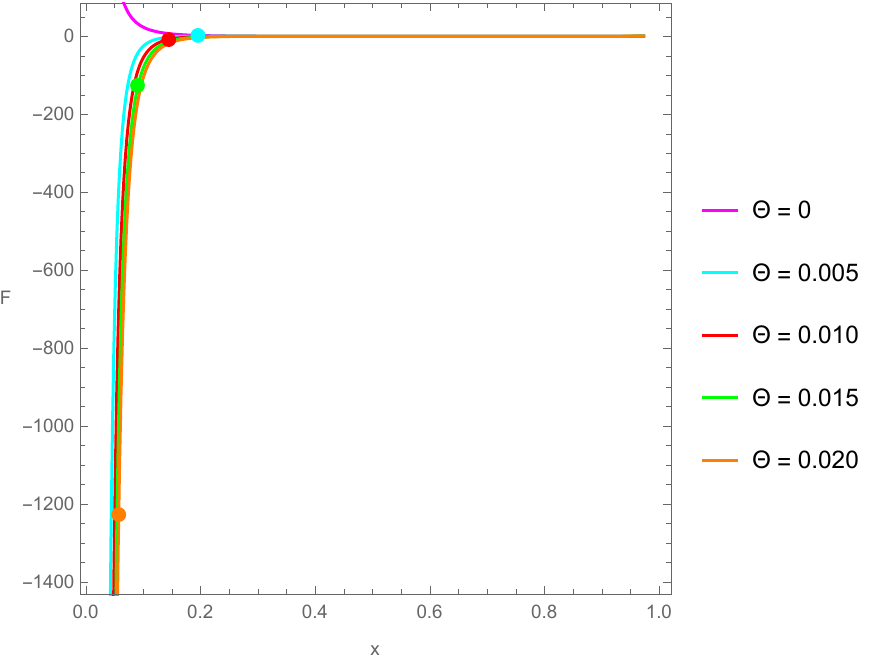}
}
\hfill
\subfloat[$F$-$x$]{%
    \includegraphics[width=0.9\linewidth]{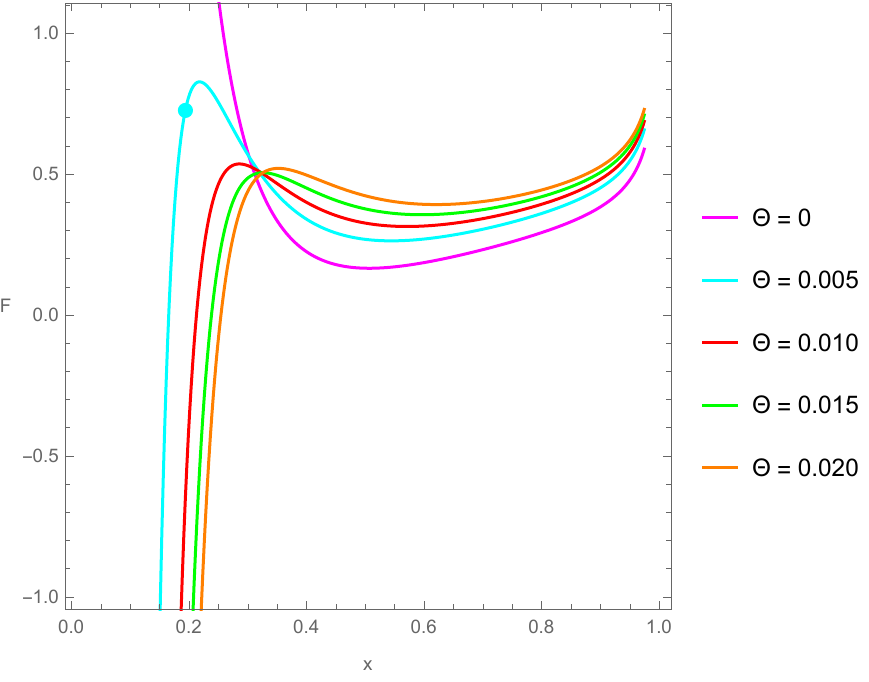}
}
\hfill
\subfloat[$F$-$T_{\mathrm{eff}}$]{%
    \includegraphics[width=0.9\linewidth]{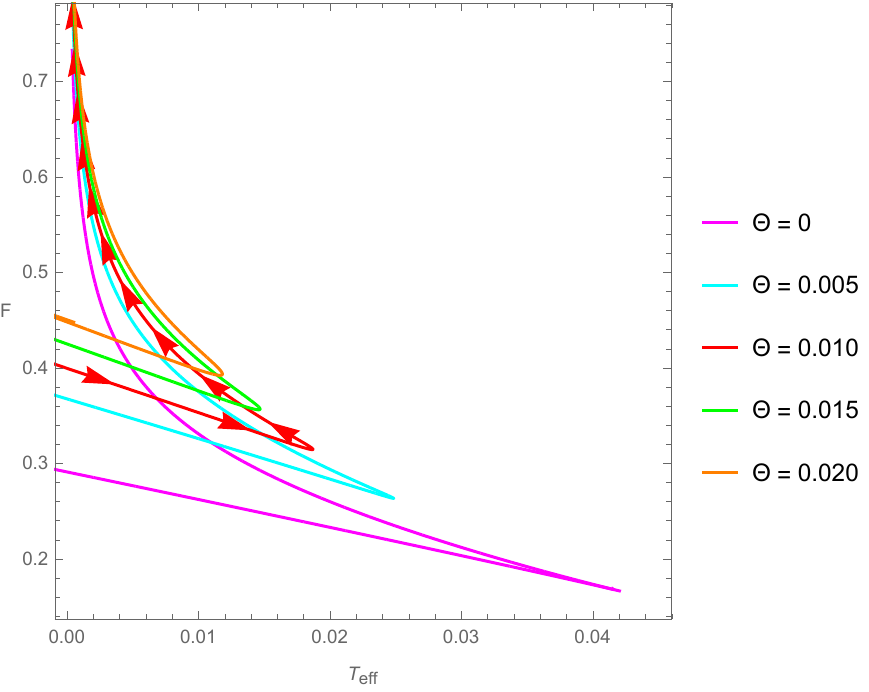}
}

\caption{
Helmholtz free energy as functions of the horizon ratio $x$ 
and the effective temperature $T_{\mathrm{eff}}$ for 
$\Theta = 0,\,0.005,\,0.010,\,0.015$ and $0.020$, 
at fixed charge $Q=0.3$ and thermodynamic volume $V=3$. 
The arrows indicate the direction of thermodynamic evolution 
with increasing $x$.
}
\label{fig:FreeEnergy}

\end{figure}

Fig.~\ref{fig:FreeEnergy} illustrates the behavior of the Helmholtz free energy $F$ as a function of the effective temperature $T_{\mathrm{eff}}$ at fixed charge $Q=0.3$ and thermodynamic volume $V=3$. The analysis is performed for the same set of noncommutative parameters $\Theta = 0,\,0.005,\,0.010,\,0.015$ and $0.020$ as those employed in the heat capacity study. For smaller black holes, characterized by lower values of the horizon ratio $x$, the free energy remains comparatively low, indicating a thermodynamically favored configuration. As $x$ increases, the free energy curve hits an inflection point, abruptly changes direction entering the  large black hole regime, the free energy rises and the system moves into a less favorable regime. The inflection point emerges at the critical temperature, which coincides with the divergence of the heat capacity and the inflection point of the entropy. This correspondence provides a consistent thermodynamic signature of the second order critical behavior of the system.

\subsection{Effective Pressure}
We now consider the behavior of the pressure–temperature ($P_{\mathrm{eff}}$–$T_{\mathrm{eff}}$) relation at fixed volume and charge. The effective pressure is defined through the Helmholtz free energy as
\begin{equation}
P_{\mathrm{eff}}=-\left(\frac{\partial F}{\partial V}\right)_{T{\mathrm{eff}},Q}.
\end{equation}
At fixed $V$, the pressure should be viewed as a function of the effective temperature alone, $P_{\mathrm{eff}}=P_{\mathrm{eff}}(T_{\mathrm{eff}})$, with its temperature dependence entirely inherited from the free energy.

Differentiating the pressure with respect to the effective temperature at fixed $V$ and $Q$ gives
\begin{equation}
\left(\frac{\partial P_{\mathrm{eff}}}{\partial T_{\mathrm{eff}}}\right)_{V,Q}
=-\left(\frac{\partial^2 F}{\partial T_{\mathrm{eff}}\partial V}\right)_{Q}.
\end{equation}

Using equation \eqref{predict}, this derivative may be written as
\begin{equation}
\left(\frac{\partial P_{\mathrm{eff}}}{\partial T_{\mathrm{eff}}}\right)_{V,Q}=-\left(\frac{\partial}{\partial V}\right)_{T_{\mathrm{eff}},Q} \left[\left(\frac{\partial F}{\partial S}\right)_{V,Q}
\frac{C_V}{T_{\mathrm{eff}}}\right].
\end{equation}

At the critical point, the heat capacity $C_V$ diverges while the effective temperature remains finite. As a consequence, the temperature derivative of the pressure becomes singular, but, remains continuous across the transition. As a result, the $P$–$T_{\mathrm{eff}}$ curve at fixed volume develops a kink or cusp at the critical temperature $T_{\mathrm{eff}}^{(c)}$.

\begin{figure}[htbp]
\centering

\subfloat[$P_{\mathrm{eff}}$- $x$]{%
    \includegraphics[width=0.9\linewidth]{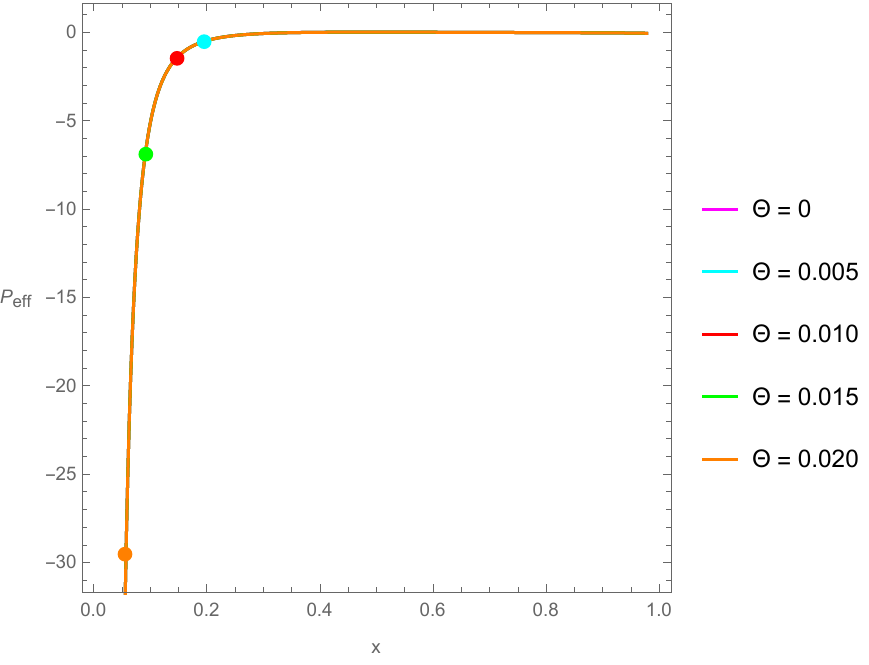}
}
\hfill
\subfloat[$P_{\mathrm{eff}}$- $x$]{%
    \includegraphics[width=0.9\linewidth]{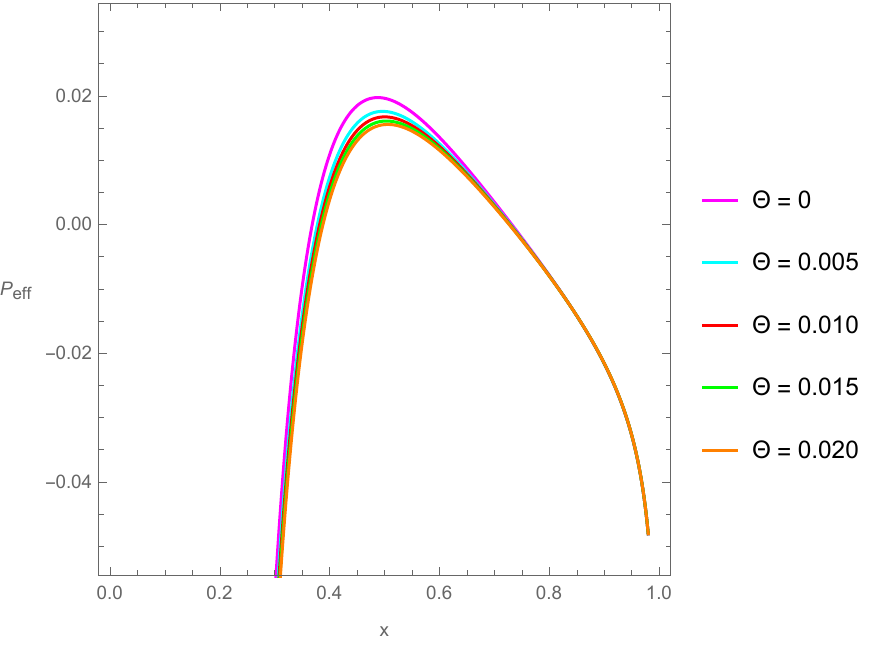}
}
\hfill
\subfloat[$P_{\mathrm{eff}}$-$T_{\mathrm{eff}}$]{    \includegraphics[width=0.9\linewidth]{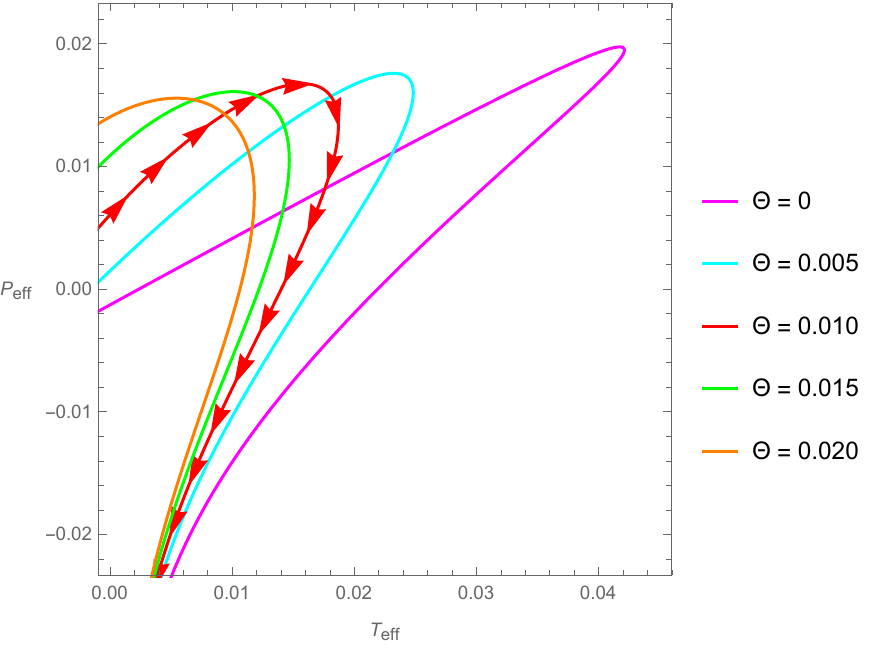}
}

\caption{
Effective pressure as functions of the horizon ratio $x$ 
and the effective temperature $T_{\mathrm{eff}}$ for 
$\Theta = 0,\,0.005,\,0.010,\,0.015$ and $0.020$, 
at fixed charge $Q=0.3$ and thermodynamic volume $V=3$. 
}
\label{fig:Pressure}

\end{figure}

Fig. \ref{fig:Pressure} illustrates the behavior of the effective pressure $P_{\mathrm{eff}}$ as a function of the effective temperature $T_{\mathrm{eff}}$ at fixed charge $Q=0.3$ and thermodynamic volume $V=3$. The analysis is carried out for the same set of noncommutative parameters $\Theta=0,0.005,0.010,0.015$ and $0.020$. The effective pressure varies smoothly with temperature. At the critical temperature, the slope of the pressure with respect to temperature becomes singular, but, remains continuous across the transition. This non-analytic behavior of $\left(\partial P_{\mathrm{eff}}/\partial T_{\mathrm{eff}}\right)_{V,Q}$ occurs at the same critical temperature where the heat capacity diverges, the entropy exhibits an inflection point and the Helmholtz free energy develops a kink.

\section{Black Hole Optics}
\subsection{Photon Dynamics}
We now investigate null geodesics in the spacetime of a Reissner–Nordström–de Sitter with non-zero noncommutative parameter, described by the metric in equation \eqref{metric}. The dynamics of a massless particle can be obtained from the Lagrangian,
\begin{align}
\mathcal{L}&=\frac{1}{2} g_{\mu\nu}\dot{x}^{\mu}\dot{x}^{\nu} \nonumber\\
&=\frac{1}{2}\left[-f(r)\dot{t}^{2}+\frac{\dot{r}^{2}}{f(r)} +r^{2}\dot{\theta}^{2}+r^{2}\sin^{2}\theta\,\dot{\phi}^{2}\right]=0,
\label{Lagrangian}
\end{align}
where the dot denotes differentiation with respect to an affine parameter.

The $t$ and $\phi$ conjugate momenta are conserved and can be identified as,
\begin{equation}
-f(r)\dot{t}=E,
\qquad
r^{2}\sin^{2}\theta\,\dot{\phi}=L,
\label{constants}
\end{equation}
where $E$ and $L$ represent the conserved photon energy and angular momentum, respectively.

Without loss of generality, the motion can be confined to the equatorial plane $\theta=\pi/2$. In this case, the null condition $\mathcal{L}=0$ reduces to,
\begin{equation}
-\frac{E^{2}}{f(r)}+\frac{\dot{r}^{2}}{f(r)}+\frac{L^{2}}{r^{2}}=0.
\label{nullcondition}
\end{equation}

Rearranging, we obtain the radial equation of motion,
\begin{equation}
\dot{r}^{2}+V_{\mathrm{eff}}(r)=E^{2},
\label{radialeq}
\end{equation}
with the effective potential
\begin{equation}
V_{\mathrm{eff}}(r)=\frac{L^{2}}{r^{2}}\,f(r).
\label{Veff}
\end{equation}

 Explicitly, the effective potential reads,
\begin{align}
V_{\mathrm{eff}}=\frac{L^2}{r^2} \left( 1 + \frac{Q^2 - 2Mr}{r^2} - \frac{4\sqrt{\Theta}\,(Q^2 - 2Mr)}{\sqrt{\pi}r^3} - \frac{\Lambda r^2}{3} \right).
\end{align}

Let us now reconsider the radial equation of motion in equation \eqref{nullcondition},
\begin{equation}
\dot r^{2} + \frac{L^{2}}{r^{2}} f(r) = E^{2}.
\end{equation}

For the study of photon trajectories, it is convenient to express the radial evolution in terms of the azimuthal coordinate $\phi$ rather than the affine parameter. Using,
\begin{equation}
\dot r = \frac{dr}{d\phi}\,\dot\phi,
\end{equation}
and recalling that in the equatorial plane $\dot\phi = L/r^{2}$, we can rewrite the radial equation in orbital form.

It is also useful to introduce the impact parameter,
\begin{equation}
b \equiv \frac{L}{E},
\end{equation}
which characterizes the asymptotic bending of photon trajectories and determines whether the photon is captured or scattered. In terms of $b$, the energy can be written as $E = L/b$.

The boundary of the black hole shadow is governed by unstable circular photon orbits (light rings). These orbits separate capture trajectories from scattering ones and define the critical impact parameter observed at infinity. Any modification of the spacetime geometry therefore leaves a direct imprint on the shadow through shifts of the light-ring radius and the associated critical impact parameter.

For null geodesics, the radial equation can be written in terms of the impact parameter $b=L/E$ as
\begin{align}
\left(\frac{dr}{d\phi}\right)^{2}
&= \frac{r^{4}}{L^{2}}\left(E^{2}-\frac{L^{2}}{r^{2}}f(r)\right) \nonumber\\
&= \frac{r^{4}}{b^{2}} - r^{2}f(r) \nonumber\\
&= r^{2} f(r) \left(\frac{r^{2}}{b^{2}f(r)} - 1 \right).
\label{orbit_equation}
\end{align}

Unstable circular photon orbits occur at extrema of the effective potential, determined by
\begin{equation}
    \frac{dV_{\text{eff}}}{dr}=0.
\end{equation}

Solving this condition gives the light-ring radius $r_c$. In the perturbative regime $\Theta \ll 1$ and $Q^{2} \ll 1$, we find
\begin{equation}
r_c = 3M - \frac{2Q^2}{3M} - \frac{16}{3\sqrt{\pi}} \sqrt{\Theta}.
\label{rc}
\end{equation}

The critical impact parameter follows from the turning-point condition $dr/d\phi=0$, yielding
\begin{equation}
b_c = \sqrt{\frac{r_c^{2}}{f(r_c)}}.
\end{equation}
Expanding consistently to the same order gives
\begin{equation}
b_c = \sqrt{3}\left(3M - \frac{Q^2}{2M}\right)
+ \frac{27\sqrt{3}}{2} M^3 \Lambda
-4 \sqrt{\frac{3}{\pi}}\,\sqrt{\Theta}.
\end{equation}

\begin{figure}[h]
    \centering
    \includegraphics[scale=0.7]{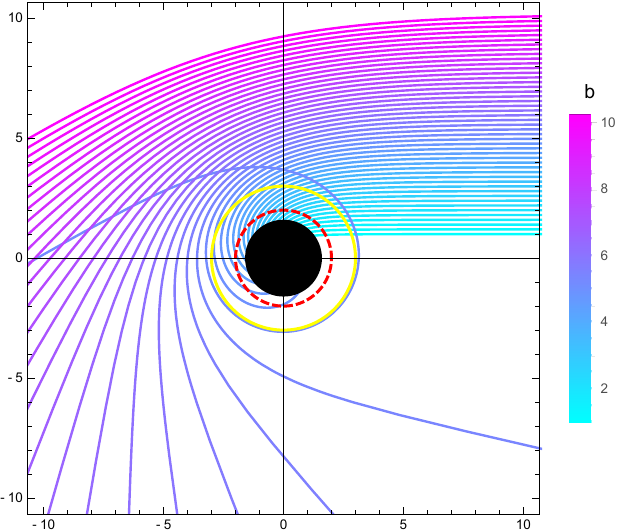}
    \includegraphics[scale=0.7]{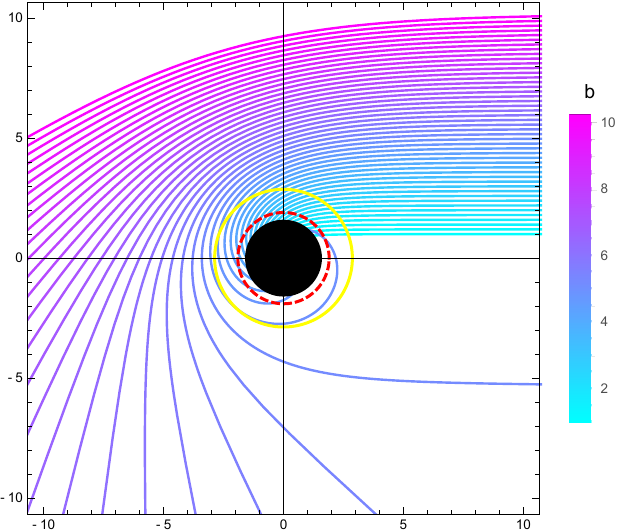}
   \caption{Ray tracing profile of a Schwarzschild-dS black hole with $M=1$ and $\Lambda=10^{-6}$ (up) and black hole with $M=1, Q=0, \Theta=0.0017$ and $\Lambda=10^{-6}$ (down).} \label{lt}
\end{figure}
\begin{figure}[h]
    \centering
     \includegraphics[scale=0.7]{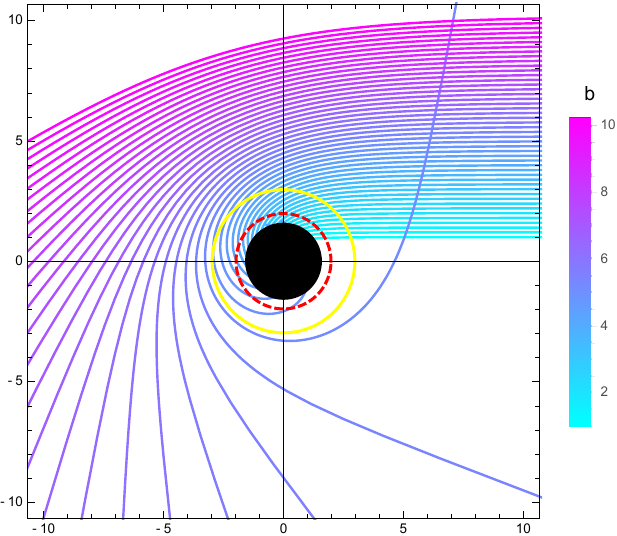}
    \includegraphics[scale=0.7]{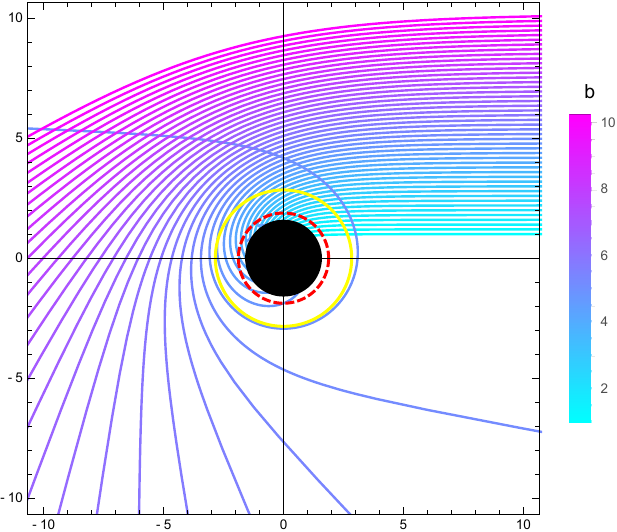}
   \caption{Ray tracing profile of a Reissner-N\"ordstrom-dS black hole with $M=1, Q=0.2$ and $\Lambda=10^{-6}$ (up) and black hole with $M=1, Q=0.2, \Theta=0.0017$ and $\Lambda=10^{-6}$ (down).} \label{lt1}
\end{figure}
The corrections induced by charge, noncommutativity and the cosmological constant affect the null geodesic behaviors as follows: The electric charge $Q^2$ and the noncommutative parameter introduces a negative shift proportional to $\sqrt{\Theta}$. In contrast, the cosmological constant $\Lambda$ does not affect the photon sphere. Any shift in the light-ring radius or in the metric function $f(r)$ propagates into the critical impact parameter. The reduction of $r_c$ due to $Q$ and $\Theta$ therefore leads to a corresponding decrease in $b_c$, implying a smaller shadow radius. The contribution from $\Lambda$ now appears through $f(r_c)$, enhancing the critical impact parameter.

Figures \ref{lt,lt1} display photon trajectories for impact parameters in the range $1 \leq b \leq 10$ in several asymptotically de Sitter black hole spacetimes. From right to left in \ref{lt} and similarly in \ref{lt1}, the panels correspond to the Schwarzschild-dS, noncommutative Schwarzschild-dS, Reissner-N\"ordstrom-dS and noncommutative Reissner-N\"ordstrom-dS black holes, respectively. In particular, the right panel of \ref{lt1} shows a charged black hole with a nonvanishing noncommutative parameter, specified by $M = 1$, $Q = 0.2$, $\Theta = 0.0017$, and $\Lambda = 10^{-6}$. The central black hole is represented by a solid black disk, while the red dashed circle indicates the event horizon at $r = 1.881$.

As the impact parameter $b$ increases, the gravitational bending of light becomes weaker and the photon trajectories gradually straighten. For small values of $b$, the gravitational field is strong enough to capture the photons, causing them to cross the event horizon. As $b$ approaches the critical value $b_c$, which separates captured and scattered trajectories, the photon asymptotically approaches the photon ring, shown as the yellow circle at $r = 2.842$. At this threshold, the trajectory winds around the unstable circular photon orbit before ultimately either escaping to infinity or falling into the black hole. For sufficiently large $b$, the gravitational field is too weak to pull the photons across the horizon and no capture occurs, all such rays are scattered. This behavior clearly demonstrates how the critical impact parameter delineates the boundary between absorption and deflection, as well as how charge and noncommutative effects modify the structure of null geodesics.

\subsection{Weak Gravitational Lensing}
We now turn to the gravitational deflection of lights in the charged black hole spacetime with non-zero commutative parameter. Gravitational lensing provides a direct probe of how the surrounding charge and matter distribution modifies the optical structure of spacetime. To quantify this effect in the weak-field regime, we adopt the Gauss-Bonnet theorem applied to the associated optical geometry, following the geometric framework developed by Gibbons and Werner \cite{Gibbons:2008rj, Waseem:2025yib}.

Unlike the conventional geodesic method, which determines the bending angle through explicit integration of the orbital equation, the Gauss-Bonnet approach encodes the deflection angle in the global topology of the optical manifold. In this formulation, gravitational lensing emerges as a purely geometric effect governed by the integrated Gaussian curvature. For alternative derivations based on elliptic integral techniques, see \cite{Ahmed:2025vww, Sucu:2025lqa}.

In the weak-field limit, the deflection angle $\alpha$ is expressed as \cite{Mandal:2023eae},
\begin{equation}
\alpha = \int_{0}^{\pi} \int_{\frac{b}{\sin \phi}}^{\infty} 
K \sqrt{g_{\mathrm{opt}}} dr d\phi, \label{alpha}
\end{equation}
where $b$ denotes the impact parameter, $K$ is the Gaussian curvature of the two-dimensional optical manifold, and $g_{\mathrm{opt}}$ represents the determinant of the optical metric.

For the static metric under consideration, the optical geometry obtained from the null condition takes the form,
\begin{equation}
dt^2 = \frac{dr^2}{f^2(r)} + \frac{r^2}{f(r)} d\phi^2,
\label{opticalmetric}
\end{equation}
from which the determinant follows as,
\begin{equation}
g_{\mathrm{opt}} = \frac{r^2}{f^3(r)}.
\end{equation}

The Gaussian curvature of this optical manifold is computed using the standard formula for a two-dimensional Riemannian space. A direct evaluation yields,
\begin{equation}
K = \frac{1}{2} \left[ \frac{1}{2} \left( \frac{df(r)}{dr} \right)^2 - f(r) \frac{d^2 f(r)}{dr^2} \right],
\end{equation}
or explicitly,
\begin{multline}
    K\approx \frac{2M}{r^{2}} + \frac{5M^{3} - 3M Q^{2}}{6 r^{2}} \Lambda \\+ \frac{2 Q^{2}-12 M^{2}}{3 \sqrt{\pi}\, r^{2}}\sqrt{\Theta}\,\Lambda+K_\Lambda,
\end{multline}
where,
\begin{equation}
    K_\Lambda=\frac{r}{3}\Lambda,
\end{equation}
is the pure de-Sitter contribution.

Here is a more natural and fluid version, while keeping a professional journal tone: It is important to note that in an asymptotically de Sitter background the concept of a deflection angle becomes more subtle. The presence of a positive cosmological constant, introduces a cosmological horizon, so the spacetime does not possess a conventional spatial infinity. As a result, light rays cannot be compared to asymptotically straight trajectories as in flat space. The deflection angle must therefore be defined in terms of finite observer and source distances by substracting the effect of the cosmological constant,
\begin{equation}
    K\to K-K_\Lambda.
\end{equation}

Integration of \eqref{alpha} then yields the deflection angle as follow,
\begin{align}
    \alpha &=\int_{0}^{\pi} \int_{\frac{b}{\sin \phi}}^{\infty} 
(K-K_\Lambda) \sqrt{g_{\mathrm{opt}}} dr d\phi\nonumber\\
&\approx \frac{1}{3 b \sqrt{\pi}} \left[ 12 M \sqrt{\pi} +  \left( 5 M^{3}  - 3 M Q^{2} \right) \sqrt{\pi}\Lambda\right.\nonumber\\\phantom{=}& \left.\phantom{=}+4\left(  Q^{2}- 6 M^{2}   \right) \sqrt{\Theta}\Lambda \right].
\end{align}

We note that the leading contribution reproduces the standard Schwarzschild term $4M/b$, i.e. in the limit where the charge and noncommutativity contribution vanishes, $Q=\Theta = 0$, the deflection angle reduces to the standard Schwarzschild expression,
\begin{equation}
\alpha = \frac{4M}{b},
\end{equation}
which is a well-known vacuum result \cite{Waseem:2025yib}.

Moreover, the electric charge introduces corrections at order $Q^2$, which reduce the bending angle relative to the neutral case. The cosmological constant contributes at higher order through mixed terms proportional to $M^3 \Lambda$ and $M Q^2 \Lambda$. Interestingly, the noncommutative parameter appears only in combination with $\Lambda$, through a term proportional to $\sqrt{\Theta}\,\Lambda$. This shows that spacetime noncommutativity does not modify the leading asymptotically flat deflection, but instead manifests through its interplay with the cosmological background. 

\section{Orbit Stability and Eikonal Resonant Scattering}
\subsection{The Lyapunov Exponent}
In this section we quantify the dynamical instability of the circular photon orbit through the Lyapunov exponent, which characterizes the exponential growth of small radial perturbations away from the photon sphere. A positive Lyapunov exponent indicates that the orbit is unstable under infinitesimal perturbations, causing photons to either fall toward the event horizon or escape toward the cosmological region. We will also show that in the eikonal regime, this instability provides a direct geometric link between null geodesic dynamics and the imaginary part of the resonant scattering's quasinormal mode frequencies.

We begin with the null Hamiltonian governing photon trajectories,
\begin{equation}
H = \mathcal{L} = \frac{1}{2} g^{\mu\nu} p_\mu p_\nu = 0 .
\end{equation}

Restricting the motion to the equatorial plane $(\theta = \pi/2)$, the Hamiltonian simplifies to
\begin{equation}
H = \frac{1}{2} \left[ - \frac{E^2}{h(r)} + h(r) p_r^2 + \frac{L^2}{r^2} \right],
\end{equation}
where $E$ and $L$ denote the conserved energy and angular momentum of the photon.

The corresponding Hamilton equations are
\begin{align}
\dot r &= h(r) p_r, \\
\dot p_r &= -\frac{1}{2}\frac{V_{\text{eff}}'(r)}{h(r)} .
\end{align}

As shown previously, a circular photon orbit at $r = r_c$ is characterized by
\begin{equation}
\dot r = 0, \qquad V_{\text{eff}}'(r_c) = 0, \qquad V_{\text{eff}}(r_c) = E^2 .
\end{equation}

At this radius the photon sits at an extremum of the effective potential, balancing gravitational attraction and centrifugal repulsion. To understand whether this orbit is stable, we slightly disturb it by introducing small radial deviations,
\begin{equation}
r = r_c + \delta r,
\qquad
p_r = \delta p_r .
\end{equation}

Keeping only terms linear in the perturbations, the equations of motion become
\begin{align}
\delta \dot r &= h(r_c)\, \delta p_r, \\
\delta \dot p_r &= -\frac{1}{2}
\frac{V_{\text{eff}}''(r_c)}{h(r_c)}\, \delta r .
\end{align}

It is convenient to express this linearized system in matrix form,
\begin{equation}
\frac{d}{d\lambda} \begin{pmatrix} \delta r \\ \delta p_r \end{pmatrix} = \begin{pmatrix} 0 & h(r_c) \\ -\dfrac{1}{2}\dfrac{V_{\text{eff}}''(r_c)}{h(r_c)} & 0 \end{pmatrix} \begin{pmatrix} \delta r \\ \delta p_r
\end{pmatrix}.
\end{equation}

The fate of the perturbations is determined by the eigenvalues of this matrix. Solving the characteristic equation,
\begin{equation}
\left| \begin{pmatrix} 0 & h(r_c) \\ -\frac{1}{2} \frac{V_{eff}''(r_c)}{h(r_c)} & 0 \end{pmatrix} - \Lambda_{LE} I \right| = 0.
\end{equation}
yields
\begin{equation}
\Lambda^2_{LE} = -\frac{V_{\text{eff}}''(r_c)}{2}.
\end{equation}

Equivalently, combining the two linearized equations gives
\begin{equation}
\delta \ddot r = -\frac{1}{2} V_{\text{eff}}''(r_c)\, \delta r ,
\end{equation}
where the sign of $V_{\text{eff}}''(r_c)$ completely controls the behavior of the perturbation.

If the circular orbit corresponds to a maximum of the effective potential,
\begin{equation}
V_{\text{eff}}''(r_c) < 0 ,
\end{equation}
the eigenvalues are real and the perturbations grow exponentially,
\begin{equation}
\delta r \sim e^{\Lambda_{LE} \lambda},
\end{equation}
in this case the photon sphere is dynamically unstable, an arbitrarily small radial displacement drives the photon away from the circular orbit.

Using
\begin{equation}
V_{\text{eff}}(r) = \frac{L^2}{r^2} h(r),
\end{equation}
the Lyapunov exponent becomes
\begin{align}
\Lambda^2_{LE} &= -\frac{L^2}{2} \frac{d^2}{dr^2} \left( \frac{h(r)}{r^2} \right)_{r=r_c} \nonumber \\ &= L^2 \left( -\frac{h''(r_c)}{2 r_c^2} + \frac{2 h'(r_c)}{r_c^3} - \frac{3 h(r_c)}{r_c^4} \right).
\end{align}

In the  regime $\Theta \ll 1$ and $Q^{2} \ll 1$, we find,
\begin{equation}
\Lambda_{LE}\approx\frac{|L|}{81 M^5} \left[ 3M \left(3M^2 + Q^2\right) + \frac{24M^2}{\sqrt{\pi}}\sqrt\Theta \right].
\end{equation}

We note that the Lyapunov exponent is real and positive. The overall $M^{-5}$ scaling suppresses the instability for large black holes, whereas the charge $Q$ increases $\Lambda_{LE}$ through the positive $Q^2$ term, thereby amplifying the orbital instability. The noncommutative parameter $\Theta$ likewise contributes positively via the $\sqrt{\Theta}$ correction, further intensifying the instability. Therefore, while increasing mass tends to stabilize the photon orbit, both charge and noncommutative effects enhance the exponential divergence.

\subsection{Eikonal Resonant Scattering}
When a black hole is perturbed, it does not settle down immediately. Instead, it undergoes a phase of damped oscillations whose frequencies are determined entirely by the geometry of the spacetime. These characteristic oscillations are known as quasinormal modes (QNMs). Their frequencies are complex: the real part sets the oscillation frequency, while the imaginary part governs the rate at which the perturbation decays. For this reason, QNMs provide a powerful probe of both the stability and dynamical response of black hole spacetimes.

The dynamics of a massless scalar field $\psi$ is described by the covariant Klein-Gordon equation,
\begin{gather}
\left[ \frac{1}{\sqrt{-g}} \partial_\mu \left( \sqrt{-g} g^{\mu \nu} \partial_\nu \right) \right] \psi = 0,
\end{gather}
where $g$ denotes the determinant of the metric tensor.

Making use of the spherical symmetry of the spacetime and the explicit metric function $h(r)$ introduced in equation \eqref{metric}, we separate variables through the standard ansatz
\begin{gather}
\psi(t,r,\theta,\phi) = e^{-i \omega t} R(r) Y_l^{m_l}(\theta,\phi),
\end{gather}
with $Y_l^{m_l}(\theta,\phi)$ representing the spherical harmonics.

Substituting this decomposition into the Klein-Gordon equation reduces the problem to a single radial equation,
\begin{gather}
\partial_r \left( h(r) r^2 \partial_r R(r) \right)
+ \left[ \omega^2 \frac{r^2}{h(r)} - l(l+1) \right] R(r) = 0,
\label{radial}
\end{gather}
which fully determines the radial behavior of the perturbation.

Introducing the tortoise coordinate $r^*$ defined by
\begin{equation}
dr^* = \frac{dr}{h(r)},
\end{equation}
and redefining the radial function according to
\begin{equation}
R(r) = \frac{1}{r} \mathcal{R}(r^*),
\end{equation}
the radial equation takes the familiar Schr\"odinger-like form
\begin{equation}
\frac{d^2 \mathcal{R}}{dr^{*2}}
+ \left[ \omega^2 - \frac{h(r)}{r^2} \left( l(l+1) + r h'(r) \right) \right]
\mathcal{R} = 0.
\end{equation}

The effective potential governing the perturbations is therefore
\begin{equation}
V_{KG}(r) = \frac{h(r)}{r^2} \left[ l(l+1) + r h'(r) \right].
\end{equation}

In the eikonal regime ($l \gg 1$), the angular momentum term dominates and the wave equation simplifies considerably,
\begin{gather}
\frac{d^2 \mathcal{R}}{dr^{*2}} + W(r)\,\mathcal{R} = 0, \\
W(r) = \omega^2 - \frac{h(r)}{r^2} l^2
= \omega^2 - \frac{l^2}{L^2} V_{\text{eff}}(r).
\end{gather}

In this high-multipole limit, the peak of the effective potential plays a central role, as it determines the dominant contribution to the quasinormal spectrum. To extract the complex frequencies, we employ the second-order WKB approximation in the Iyer-Will formulation \cite{Schutz:1985km}. The quantization condition reads
\begin{equation}
\frac{W(r_0)}{\sqrt{2 W^{(2)}(r_0)}} = -i \left( n + \frac{1}{2} \right), \label{WKB1}
\end{equation}
where $n = 0,1,2,\dots$ labels the overtone number and
\begin{equation}
W^{(2)}(r_0) = \left. \frac{d^2 W}{d {r^*}^2} \right|_{r=r_0}.
\end{equation}

The radius $r_0$ corresponds to the maximum of the effective potential and coincides with the circular null orbit, $r_0 = r_c$, namely the photon sphere.

Applying the WKB quantization leads to
\begin{multline}
\omega_{\text{QNM}} \approx \frac{l}{|L|} \sqrt{V_{\text{eff}}(r_c)} \\- i \left( n + \frac{1}{2} \right) \sqrt{- \frac{1}{2 V_{\text{eff}}(r_c)} \left. \frac{d^2 V_{\text{eff}}}{d {r^*}^2} \right|_{r=r_c} },
\end{multline}
where we can use the chain rule of derivative to obtain
\begin{equation}
\left. \frac{d^2 V_{\text{eff}}}{d {r^*}^2} \right|_{r=r_c} = h^2(r_c) V_{\text{eff}}''(r_c),
\end{equation}
and the frequency takes the compact form
\begin{equation}
\omega_{\text{QNM}} = \frac{l}{|L|}\sqrt{V_{\text{eff}}(r_c)} - i\left(n+\frac{1}{2}\right) \frac{|\Lambda|}{L^{2}} r_c^{2}\sqrt{V_{\text{eff}}(r_c)}.
\end{equation}

We note that the real part of the quasinormal frequency is governed by the height of the effective potential at the photon sphere and reflects the orbital motion of null geodesics. The imaginary part, on the other hand, is controlled by the curvature of the potential at that same location. It therefore measures how rapidly nearby photon trajectories diverge, which is precisely quantified by the Lyapunov exponent $\Lambda_{LE}$. Thus, the damping rate of quasinormal modes is directly determined by the instability timescale of the circular null orbit. Any modification of the metric function $h(r)$ arising from charge, cosmological constant or noncommutative smearing therefore is expeced to leave an imprint on both the oscillation frequency and the decay rate of the spectrum.

Let us consider the special case $Q^2\ll 1$ and $\Theta\ll 1$, a series expansion yields an analytical expression of the eikonal QNM in noncommutative RN-dS black hole spacetime,
\begin{multline}
\omega_{QNM}= \frac{l}{18 M^3 \sqrt{3}} \Bigg[ 6M^2+Q^2 -27M^4\Lambda  + \frac8{\sqrt\pi}\sqrt\Theta M \Bigg]\\- i\frac{(2n+1)}{54\,M^3\sqrt{3}}\Bigg[
\sqrt18M^2+Q^2- 81M^4\Lambda+\frac8{\sqrt\pi}\sqrt\Theta M \Bigg]. \label{QNMApprox}
\end{multline}

Now, let us consider the imaginary part of $\omega_{QNM}$, which controls the decay timescale of the perturbation. The factor $M^{-3}$ indicates that increasing the black hole mass suppresses the magnitude of the imaginary part, leading to longer-lived modes. The electric charge $Q$ enters with a positive contribution inside the bracket, enhancing the absolute value of the imaginary part and therefore increasing the damping rate. In contrast, the cosmological constant $\Lambda$ appears with a negative sign through the $-81M^4\Lambda$ term, partially reducing the magnitude of the damping and thus slightly prolonging the lifetime of the mode. Finally, the noncommutative parameter $\Theta$ strengthening the decay rate via the $8\sqrt{\Theta}M$ term. 

As a final consistency check, let us recover the Schwarzschild limit. For a pure Schwarzschild black hole, the effective potential governing null geodesics is
\begin{equation}
V_{\text{eff}}(r)=\frac{L^{2}}{r^{2}}\left(1-\frac{2M}{r}\right),
\end{equation}
with the unstable circular null orbit located at the photon sphere $r_c = 3M$. At this radius the potential reaches its maximum, and small radial perturbations grow exponentially. Evaluating the relevant quantities at $r_c$ yields
\begin{equation}
\Lambda = \frac{|L|}{9 M^{2}},
\qquad
V_{\text{eff}}(r_c)=\frac{L^{2}}{27 M^{2}}.
\end{equation}

Substituting these into the general eikonal expression immediately reproduces the well-known result \cite{Ferrari:1984zz},
\begin{equation}
\omega_{QNM}^{\rm Schwarzschild} = \frac{l}{3\sqrt{3} M} - \frac{i\left(n+\frac{1}{2}\right)}{3\sqrt{3} M}.
\end{equation}

\section{Conclusion}
In this work, we have investigated Reissner–Nordström–de Sitter black holes in a noncommutative geometric framework, where the point-like mass is replaced by a Lorentzian-smeared matter distribution characterized by a minimal length scale. The presence of a positive cosmological constant generates three horizons; however, since the inner (Cauchy) horizon does not directly participate in the thermodynamic description, our analysis focuses on the event and cosmological horizons. A central difficulty in de Sitter thermodynamics arises from the fact that these two horizons generally possess different temperatures, preventing global thermal equilibrium.

To overcome this issue, we employed an effective thermodynamic framework based on the extended Iyer–Wald formalism, allowing the two-horizon system to be described consistently in terms of global effective quantities. A key element of this construction is a modification of the entropy, ensuring that the effective temperature correctly reproduces the lukewarm configuration, in which the two horizon temperatures coincide. We demonstrated that, in the commutative limit, this entropy correction reduces to previously established results. When noncommutativity is incorporated, additional contributions emerge, and notably, the modified entropy admits a closed analytic form. From this corrected entropy, we derived the remaining thermodynamic variables via the effective first law. The thermodynamic volume naturally reflects the geometric separation between the event and cosmological horizons, completing a self-consistent effective description of the noncommutative RN–dS spacetime.

We then examined both local and global thermodynamic stability in the canonical ensemble by analyzing the entropy, heat capacity, Helmholtz free energy, and effective pressure. The heat capacity serves as the primary diagnostic of local stability. Our investigation reveals two distinct thermodynamic branches separated by a divergence of the heat capacity, signaling a critical point. The small black hole branch contains a region of positive heat capacity and is therefore locally stable, whereas the large black hole branch exhibits negative heat capacity and is unstable. The inclusion of the noncommutative parameter reduces the extent of the stable region and enhances the instability of the large black hole branch, indicating that minimal-length effects substantially reshape the stability structure.

The remaining thermodynamic quantities consistently support this picture. The entropy remains continuous across the transition but develops an inflection point at the critical temperature. The Helmholtz free energy is also continuous, while its second temperature derivative becomes singular, identifying the transition as second order in the Ehrenfest classification. A corresponding cusp structure in the free-energy diagram provides a clear geometric signature of this critical behavior. Moreover, the effective pressure exhibits non-analytic behavior at the same critical temperature, with a divergent temperature derivative but no discontinuity. The simultaneous appearance of these features confirms the internal consistency of the effective thermodynamic framework.

We also explored the optical properties of the noncommutative RN–dS spacetime. Starting from the photon effective potential, we derived the orbital equation in terms of the impact parameter and determined the critical impact parameter that separates capture from scattering trajectories. Consistent with the behavior of the light ring, both the electric charge and the noncommutative parameter reduce the photon-sphere radius, leading to a corresponding decrease in the critical impact parameter. The cosmological constant modifies the global structure of the spacetime and indirectly enlarges the capture region through its influence on the effective potential.

In the weak-field regime, we computed the gravitational deflection angle using the Gauss–Bonnet method. After subtracting the pure de Sitter contribution, we obtained a closed analytic expression for the bending angle. The leading term reproduces the Schwarzschild result, providing a nontrivial consistency check. Electric charge introduces corrections that reduce the bending relative to the neutral case. Interestingly, the noncommutative parameter appears only in combination with the cosmological constant, indicating that its optical influence is intrinsically tied to the cosmological background rather than to the asymptotically flat sector.

Finally, we investigated the dynamical response of the system through the Lyapunov exponent and the imaginary part of the quasinormal mode frequencies. This analysis makes explicit the correspondence between the instability of photon orbits and the damping of perturbations. Increasing the black hole mass suppresses the Lyapunov exponent and reduces the magnitude of the imaginary part of $\omega_{QNM}$, leading to weaker orbital instability and longer-lived modes. In contrast, electric charge enhances both the geodesic instability and the damping rate. The cosmological constant partially reduces the damping, slightly prolonging the lifetime of perturbations, whereas the noncommutative parameter increases both the Lyapunov exponent and the decay rate.

\newpage
\appendix

\newcommand{\uRT}{u} 
\begin{widetext}
\section{Effective Pressure}
\label{app:P_full}
For the sake of simplicity, we introduce,
\begin{align}
\uRT &\equiv (x^{3}-1)^{1/3},\qquad \uRT^{2}=(x^{3}-1)^{2/3}, \\[0.4em] P(r_c,x,Q,\Theta) &= \mathcal{K}(x) \Big[ r_c^{2}x^{2}\,A(x;\uRT)+Q^{2}\,B(x;\uRT) \nonumber\\ &\hspace{7em} +2\sqrt{\Theta}\Big(-r_c^{2}x^{2}\,C(x;\uRT)+Q^{2}\,D(x;\uRT)\Big) \Big],
\label{eq:P_factorised}
\end{align}
where the four auxiliary functions are given by
\begin{align}
K(x;\uRT)&\equiv  \frac{7}{280\,\pi\,r_c^{4}}\; \frac{1}{(1-x)\,x^{3}\,(1+x+x^{2})^{2}\,(1+x^{4})\,\uRT} \\[0.3em] A(x;\uRT)&\equiv -8-8x^{10}-3\uRT +x^{8}\big(16-3\uRT\big) -8x^{4}(-1+\uRT) +2x^{3}(4+\uRT) \nonumber\\ &\quad -8x^{5}(4+\uRT) +2x^{6}(4+\uRT) -x(8+3\uRT) \nonumber\\ &\quad +2x^{2}(8+3\uRT) -x^{9}(8+3\uRT) +x^{7}(8+6\uRT), \\[0.3em] B(x;\uRT)&\equiv 8+8x^{12}+3\uRT +16x^{6}(3+\uRT) -2x^{4}(4+\uRT) -2x^{7}(4+\uRT) +8x^{5}(-1+2\uRT) \nonumber\\ &\quad +x^{9}(-32+3\uRT) +x(8+3\uRT)+x^{2}(8+3\uRT) \nonumber\\ &\quad +x^{10}(8+3\uRT)+x^{11}(8+3\uRT) -x^{8}(8+11\uRT)-x^{3}(32+11\uRT),\\C(x;\uRT)&\equiv -112x^{3}\uRT(3+\uRT) +14\uRT(8+3\uRT) +x^{11}\uRT(224+39\uRT) +x^{13}(115+112\uRT) \nonumber\\ &\quad +x^{8}\big(230-224\uRT-262\uRT^{2}\big) +x^{2}\big(230+224\uRT-6\uRT^{2}\big) -2x^{10}\big(115+168\uRT+3\uRT^{2}\big) \nonumber\\ &\quad -4x^{9}\big(115+140\uRT+28\uRT^{2}\big) +x\big(115+224\uRT+39\uRT^{2}\big) +x^{12}\big(230+224\uRT+42\uRT^{2}\big) \nonumber\\ &\quad +2x^{6}\big(115+280\uRT+74\uRT^{2}\big) -2x^{4}\big(115+280\uRT+131\uRT^{2}\big) \nonumber\\ &\quad +x^{5}\big(-460-224\uRT+141\uRT^{2}\big) +x^{7}\big(230+560\uRT+141\uRT^{2}\big), \\[0.3em]
\end{align}
\begin{align}
     D(x;\uRT)&\equiv 14\uRT(8+3\uRT) +x^{15}(115+112\uRT) +x^{9}\big(460-112\uRT-405\uRT^{2}\big) +x^{3}\big(345-121\uRT^{2}\big) \nonumber\\ &\quad +x^{12}\big(-230+36\uRT^{2}\big) +x^{2}\big(230+336\uRT+36\uRT^{2}\big) +x\big(115+224\uRT+39\uRT^{2}\big) \nonumber\\ &\quad +x^{13}\big(345+336\uRT+39\uRT^{2}\big) +x^{14}\big(230+224\uRT+42\uRT^{2}\big) \nonumber\\ &\quad +4x^{8}\big(115+196\uRT+52\uRT^{2}\big) -2x^{4}\big(115+280\uRT+61\uRT^{2}\big) \nonumber\\ &\quad -2x^{10}\big(345+392\uRT+61\uRT^{2}\big) +2x^{7}\big(230+392\uRT+71\uRT^{2}\big) \nonumber\\ &\quad +2x^{6}\big(-345-56\uRT+104\uRT^{2}\big) -x^{11}\big(460+560\uRT+121\uRT^{2}\big) \nonumber\\ &\quad -x^{5}\big(460+784\uRT+405\uRT^{2}\big).
\end{align}
\end{widetext}

\begin{widetext}
\section{Heat Capacity}\label{app:Heat}
The explicit expression of the heat capacity is presenten as follows,
\begin{equation}
C_{0} = -\,\frac{ 2 \pi r_c^{2} x^{2} (1+x^{4})^{2} \Big[ r_c^{2} x^{2}\,\mathcal{N}_{1}(x) - Q^{2}\,\mathcal{N}_{2}(x) \Big] }{ (x^{2}-1)(1+x+x^{2}) \Big[ r_c^{2} x^{2}\,\mathcal{D}_{1}(x) - Q^{2}\,\mathcal{D}_{2}(x) \Big] },
\end{equation}

\begin{equation}
C_{1/2} = \frac{ 16 \sqrt{\pi}r_c x (1+x^{4})^{2} \Big[ r_c^{4} x^{4}\,\mathcal{P}_{1}(x) - Q^{2} r_c^{2} x^{2}\,\mathcal{P}_{2}(x) + 2 Q^{4}\,\mathcal{P}_{3}(x) \Big] }{ (x-1)(1+x)^{2}(1+x+x^{2})^{2} \Big[ r_c^{2} x^{2}\,\mathcal{D}(x) - Q^{2}\,\mathcal{E}(x) \Big]^{2} }.
\end{equation}
where,
\begin{equation}
\mathcal{N}_{1}(x)=1+x-2x^{2}+x^{3}+x^{4},
\end{equation}

\begin{equation}
\mathcal{N}_{2}(x)=1+x+x^{2}-2x^{3}+x^{4}+x^{5}+x^{6}.
\end{equation}

\begin{equation}
\mathcal{D}_{1}(x)=1+4x^{2}-6x^{3}+9x^{4}-8x^{5}+9x^{6}-6x^{7}+4x^{8}+x^{10},
\end{equation}

\begin{equation}
\mathcal{D}_{2}(x)=3+3x^{2}+7x^{4}-2x^{6}+7x^{8}+3x^{10}+3x^{12}.
\end{equation}

\begin{equation}
\begin{aligned}
\mathcal{P}_{1}(x) &= 1 + 3x + 2x^{2} + 4x^{3} + 4x^{4} + 10x^{5} + 4x^{7} + 9x^{8} + 6x^{9} + 9x^{10} \\ &\quad + 4x^{11} + 10x^{13} + 4x^{14} + 4x^{15} + 2x^{16} + 3x^{17} + x^{18}.
\end{aligned}
\end{equation}

\begin{equation}
\begin{aligned}
\mathcal{P}_{2}(x) &= 3 + 9x + 9x^{2} + 12x^{3} + 21x^{4} + 27x^{5} + 24x^{6} + 9x^{7} \\ &\quad + 26x^{8} + 31x^{9} + 30x^{10} + 31x^{11} + 26x^{12} + 9x^{13} + 24x^{14} \\ &\quad + 27x^{15} + 21x^{16} + 12x^{17} + 9x^{18} + 9x^{19} + 3x^{20}.
\end{aligned}
\end{equation}

\begin{equation}
\begin{aligned}
\mathcal{P}_{3}(x) &= 1 + 3x + 6x^{2} + 5x^{3} + 8x^{4} + 12x^{5} + 17x^{6} + 11x^{7} + 7x^{8} \\ &\quad + 11x^{9} + 19x^{10} + 20x^{11} + 19x^{12} + 11x^{13} + 7x^{14} + 11x^{15} \\ &\quad + 17x^{16} + 12x^{17} + 8x^{18} + 5x^{19} + 6x^{20} + 3x^{21} + x^{22}.
\end{aligned}
\end{equation}

\begin{equation}
\mathcal{D}(x) = 1 + 4x^{2} - 6x^{3} + 9x^{4} - 8x^{5} + 9x^{6} - 6x^{7} + 4x^{8} + x^{10},
\end{equation}

\begin{equation}
\mathcal{E}(x) = 3 + 3x^{2} + 7x^{4} - 2x^{6} + 7x^{8} + 3x^{10} + 3x^{12}.
\end{equation}

\end{widetext}

\providecommand{\noopsort}[1]{}\providecommand{\singleletter}[1]{#1}%

\end{document}